\documentclass[aps,prd,showpacs,superscriptaddress,nofootinbib,notitlepage,preprintnumbers,reprint]{revtex4-1}
\pdfoutput=1
\usepackage[T1]{fontenc}
\usepackage{bm,amsmath,amssymb,color,bbm,slashed,MnSymbol}
\usepackage[pdftex]{graphicx}
\usepackage[colorlinks=true, pdfstartview=FitH, linkcolor=blue, citecolor=red, urlcolor=blue]{hyperref}

\DeclareMathOperator\diag{diag}
\newcommand{\ee}{e}
\DeclareMathOperator\tr{tr}

\newcommand{\der}{\partial}
\renewcommand{\bar}[1]{\overline{#1}}
\newcommand{\calO}{\mathcal{O}}
\newcommand{\dd}{\mathrm{d}}
\newcommand{\bep}{\begin{pmatrix}} 
\newcommand{\eep}{\end{pmatrix}}

\newcommand{\SU}{\text{SU}}
\newcommand{\SO}{\text{SO}}

\newcommand{\U}{\text{U}}
\newcommand{\1}{\mathbbm{1}}
\newcommand{\RR}{\mathbb{R}}

\newcommand{\ZZ}{\mathbb{Z}}
\renewcommand{\epsilon}{\varepsilon}
\newcommand{\wt}[1]{\widetilde{#1}}

\def\ba#1\ea{\begin{align}#1\end{align}}

\def\mkakko#1{\left( #1 \right)}
\def\ckakko#1{\left\{ #1 \right\}}
\def\kkakko#1{\left[ #1 \right]}

\newcommand{\DD}{\slashed{D}}
\newcommand{\muu}{\hat{\mu}}
\newcommand{\mm}{\hat{m}}
\renewcommand{\aa}{\hat{a}}

\newcommand{\QCD}{\text{QCD}}
\newcommand{\NfD}{N_f^{\rm D}}
\newcommand{\Nc}{N_c}
\newcommand{\KK}{\text{KK}}
\newcommand{\rma}{\mathtt{a}}
\newcommand{\rmb}{\mathtt{b}}
\newcommand{\rmA}{\textsf{A}}
\newcommand{\rmB}{\textsf{B}}
\begin{document}
\preprint{RIKEN-QHP-256} 
\title{
Phases of circle-compactified QCD with adjoint fermions at finite density
}

\author{Takuya Kanazawa} 
\affiliation{iTHES Research Group and Quantum Hadron Physics Laboratory, 
RIKEN, 6-7-1 Minatojima-minamimachi, Chuo-ku, Kobe, Hyogo 650-0047, Japan}
\author{Mithat \"Unsal}
\affiliation{Department of Physics, North Carolina State University, Raleigh, North Carolina 27695, USA}
\author{Naoki Yamamoto}
\affiliation{Department of Physics, Keio University, Yokohama 223-8522, Japan}
\allowdisplaybreaks
\begin{abstract}
	We study chemical-potential dependence of confinement and mass gap 
	in  QCD with adjoint fermions in spacetime with one spatial compact direction. By calculating 
	the one-loop effective potential for the Wilson line  in the presence of chemical potential, 
	we show that a center-symmetric phase and a center-broken phase alternate when the chemical 
	potential in unit of the compactification scale is increased. In the center-symmetric phase 
	we use semiclassical methods to show that photons in the magnetic bion plasma acquire 
	a mass gap that grows with the chemical potential as a result of anisotropic 
	interactions between monopole-instantons. 
	For the neutral fermionic sector which remains gapless perturbatively, 
	there are two possibilities at non-perturbative level.  Either to remain 
	gapless (unbroken global symmetry), or to  undergo a novel superfluid transition through a 
	four-fermion interaction (broken global symmetry). If the latter is the case,  
	 this leads  to an  energy gap of quarks proportional to  a new  nonperturbative scale 
	$L^{-1}\exp[-1/(g^4 \mu L)]$, where $L$ denotes the circumference of $S^1$,   the low-energy 
	is described as a Nambu--Goldstone mode associated with the baryon number, and 
	there exists a new type of BEC-BCS crossover of the diquark pairing as a function of the 
	compactification scale at small chemical potential.  
\end{abstract}
\maketitle

\section{Introduction}

Semiclassical analysis is a powerful tool to study quantum systems 
nonperturbatively. In QCD, instantons have long been a subject of 
intensive research \cite{Gross:1980br,Schafer:1996wv,
Diakonov:1995ea,*Diakonov:2002fq,Vandoren:2008xg}.  
They played crucial roles in phenomenological models  of  
spontaneous chiral symmetry breaking,  as well as the U(1) problem and the strong CP problem.  
 In dense quark matter, instanton effects are semiclassically reliable as it has been  
 been widely recognized \cite{Shuryak:1982hk,Rapp:1997zu,Alford:1997zt,
Carter:1998ji,Rapp:1999qa,Son:2001jm,Schafer:1998up,*Schafer:2002ty}. 

Our understanding of semiclassical aspects of non-Abelian gauge theories including QCD 
was significantly advanced with two related progress:    
\begin{itemize}
\item KvBLL monopole-instantons on  $\RR^3 \times S^1$
\cite{Lee:1997vp, Lee:1998bb, Kraan:1998pm, Kraan:1998sn}, 
{\item Semiclassically calculable domains  on $\mathbb R^3 \times S^1$ in (non-supersymmetric) QCD-like and Yang-Mills theories 
\cite{Unsal:2007vu, Unsal:2007jx,Shifman:2008ja, Unsal:2007jx,Argyres:2012ka}.}
\end{itemize}
Monopole-instantons  carry  fractional topological charge, and 
obey the Nye--Singer index theorem \cite{Nye:2000eg, Poppitz:2008hr} (which is a refinement of the Atiyah--Singer index). 
The sum of the fermion zero modes of the monopole-instantons is equal to the one of instanton in four dimensions (4d), hence, they are more relevant 
for chiral symmetry breaking.  In sharp contrast to 4d instantons, monopole-instantons on  $\mathbb R^3 \times S^1$ explicitly    depend 
on the background holonomy (Wilson line)  field. Therefore, monopole-instantons  provide a concrete 
connection between  center-symmetry realization (confinement in a thermal context)  and chiral symmetry realization.

In ${\cal N}=1$ supersymmetric Yang-Mills (SYM)  theory on $S^1\times \RR^3$, the ensemble of 
monopole-instantons  generates  the gluino condensate 
\cite{Davies:1999uw,*Davies:2000nw}. The  applicability of semi-classics to non-supersymmetric 
QCD-like theories was shown 
in Refs.~\cite{Unsal:2007vu,*Unsal:2007jx,Shifman:2008ja,
Unsal:2008ch,Meisinger:2009ne,Argyres:2012ka,Ogilvie:2012is}. (Also see Ref.~\cite{Diakonov:2004jn}.) 
In such theories gauge symmetry ``breaking'' occurs  
at small $S^1$ (at weak coupling) due to the Hosotani mechanism 
\cite{Hosotani:1983xw,*Hosotani:1988bm}. 
In SYM as well as in adjoint QCD, in the Abelian regime, 
topologically neutral molecules of monopoles called 
 \emph{magnetic bions} form and generate a bosonic potential  
that engenders a mass gap of gluons and confines quarks  
\cite{Unsal:2007vu,*Unsal:2007jx,Argyres:2012ka}. 

In this work, we study phases of compactified adjoint QCD 
on $\mathbb R^3 \times S^1$ at nonzero quark chemical potential $\mu$. 
It should be noted that $S^1$ in this 
paper is a compactified \emph{spatial} direction along which fermions obey 
the periodic boundary condition (PBC); the imaginary-time direction is infinite 
and the system is at zero temperature. 
This setup must not be confused with previous studies of a holonomy potential 
for \emph{thermal} $S^1$ at $\mu\ne0$ 
\cite{KorthalsAltes:1999cp,Myers:2009df,Hands:2010zp,Bruckmann:2013rpa,Liu:2016thw}.

This theory is semiclassically calculable at small  $S^1$ for any value of 
$\mu$.%
\footnote{At large $\mu$  and for   values of   $S^1$ larger than the strong length 
scale, there are sub-sectors of the theory  amenable to the weak coupling treatment, 
however, there are sectors of the theory which are still strongly coupled, in particular, 
the gauge sector.} 
For generic colors $(N_c)$ and Dirac flavors $(\NfD)$ 
we compute the perturbative one-loop  potential for the Wilson line holonomy. 
We find that turning on a small $\mu$ weakens gauge symmetry 
breaking, while larger $\mu$ induces an infinite number of oscillatory transitions 
between a gauge-symmetry-restored phase and a broken phase.  
In the specific case 
of $N_c=2$ and $\NfD=1$, we  show that perturbatively induced 
four-fermion operators%
\footnote{The monopole operators which also possess four fermionic zero modes 
are nonperturbatively weak in the semiclassical domain.}  
that have so far been neglected in studies of the Hosotani phase can play a pivotal role 
in generating diquark condensation and drive the system into a superfluid phase 
with broken $\U(1)_B$ baryon number symmetry.   
Our calculations are performed in a theoretically controlled setting thanks to the 
small running coupling and complete Abelianization of the gauge group at small $S^1$.
This is a unique opportunity in which one can derive both continuous symmetry breaking 
and confinement from first principles analytically, providing a valuable laboratory to 
study the strongly coupled QCD vacuum. 

As a quick guide to readers, we summarize a cascade of 
symmetry breaking in the presence of $\mu$ for $N_c=2$ and $\NfD=1$ 
when $\mu$ is small compared to the compactification scale $L^{-1}$:  
\ba
	& \U(2)_f \times [\SU(2)]_c 
	\label{eq:sym1}
	\\
	\xrightarrow{\text{anomaly}} ~ & \big((\ZZ_8)_A \times \SU(2)_f \big) /\ZZ_2 \times [\SU(2)]_c
	\label{eq:sym2}
	\\
	\xrightarrow{~\mu\ne 0~} ~ & \big((\ZZ_8)_A \times \U(1)_B \big) /\ZZ_2 \times [\SU(2)]_c
	\label{eq:sym3}
	\\
	\xrightarrow{~\text{Adj.~Higgs}~} ~ & \big((\ZZ_8)_A \times \U(1)_B \big) /\ZZ_2 \times [\U(1)]_c
	\label{eq:sym4}
	\\
	\xrightarrow{~\langle \chi \rangle = 0,\,\pi~}~ & \big( (\ZZ_4)_A \times \U(1)_B \big)/\ZZ_2 \times [\U(1)]_c
	\label{eq:sym5}
	\\
	\xrightarrow{~\langle \psi\psi \rangle\ne 0~} ~ & (\ZZ_2)_R \times (\ZZ_2)_L \times [\U(1)]_c  
	\label{eq:sym6}
\ea
Here $[...]_c$ denotes the unbroken part of the gauge group and $\chi$ is the dual photon field. 
In \eqref{eq:sym2}--\eqref{eq:sym5} we have factored a common $\ZZ_2$ group to avoid double counting. 
Detailed explanations will follow later.  

\begin{figure}[tb]
	\centering
	\includegraphics[width=.37\textwidth]{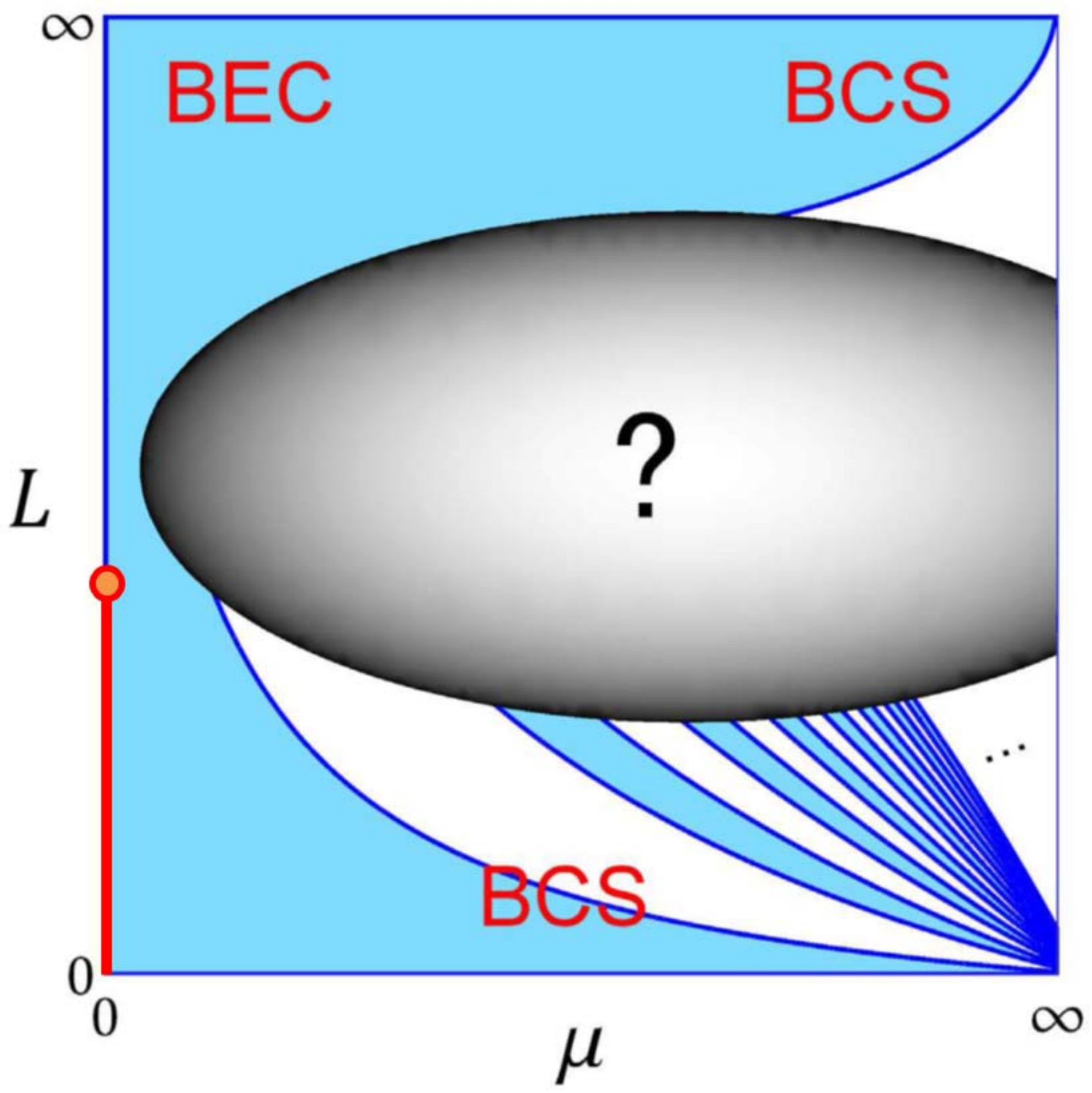}
	\vspace{-2mm}
	\caption{\label{fg:pd}%
	Phase diagram of $\Nc=2$ adjoint QCD on $\RR^3\times S^1$ 
	in the chiral limit based on the result of our investigation, with $L$ the circumference of $S^1$.  
	The blue region represents the center-symmetric phase and the white 
	region the center-broken phase. The diquark condensate $\langle\psi\psi\rangle$ 
	vanishes on the $L\lesssim \Lambda_{\QCD}^{-1}$ part of the $L$ axis 
	(represented by a red line with a critical end point). 
	The superfluid phase at weak (strong) coupling 
	is indicated by the label ``BCS'' (``BEC''), respectively. 
	The phase structure at intermediate $L$ is currently unknown. 
	}
\end{figure}

In Fig.~\ref{fg:pd} we present a schematic phase diagram of adjoint QCD 
with the aim to illustrate how our analysis in this paper in the small-$S^1$ regime 
should fit in the global phase diagram from small to large $S^1$.  
Adjoint QCD on $\RR^4$ is believed to exhibit a relativistic analogue of 
the BEC-BCS crossover \cite{Chen20051,ZwergerBook2012} 
when the chemical potential is swept across $\Lambda_{\QCD}$ 
\cite{Kogut:2000ek,Hands:2000ei,Splittorff:2000mm,Hands:2001ee,
Nishida:2005ds,Sun:2007fc,He:2010nb,Kanazawa:2011tt,Kanazawa:PhD}: 
\begin{itemize}
	\item[$\checkmark$] 
	At $\mu=0$ and large $L$, the center symmetry is preserved, while the flavor symmetry is believed to be 
	spontaneously broken as $\SU(2)_f \to \SO(2)_f$ for $\NfD=1$ \cite{Peskin:1980gc}. 
	The physical spectrum contains a light diquark that undergoes a Bose condensation for $\mu>m_{\pi}/2$ 
	and breaks $\U(1)_B$. This onset of superfluidity is well described by 
	chiral perturbation theory \cite{Kogut:2000ek}.  
	\item[$\checkmark$] 
	In the asymptotic region $\mu\gg\Lambda_{\QCD}$, on $\RR^4$, 
	quarks acquire a large energy gap $\Delta\sim \mu\,\ee^{-1/g}$ from the perturbative one-gluon exchange 
	at the Fermi surface \cite{Son:1998uk,Schafer:1999jg} and the 
	low-energy effective theory is anisotropic Yang-Mills theory with a confining scale $\Lambda'$  
	that goes down to zero as $\mu\to \infty$ due to medium effects \cite{Rischke:2000cn}. 
	The compactification $L<\infty$ acts on gluons as a thermal scale and triggers a 
	deconfinement transition at $L^{-1}\sim \Lambda'$. This is why the phase transition line 
	must emanate from the top-right corner of the phase diagram in Fig.~\ref{fg:pd}.  
	\item[$\checkmark$] 
	At $\mu=0$, if $L\Lambda_{\QCD}\ll1$, the system enters a center-symmetric 
	weakly-coupled regime where the discrete chiral symmetry remains 
	broken, while the continuous  chiral symmetry is restored 
	\cite{Unsal:2007vu}. This is indicated by a red line 
	in the bottom-left corner of Fig.~\ref{fg:pd}.  
	\item[\large\textcircled{\normalsize $\checkmark$}]
	The above three domains are well understood by now. In this paper, we venture into the novel 
	regime of small $S^1$ and any $\mu$. Among other things we uncover, in addition to a series of 
	confinement--deconfinement transitions, we find  that a novel mechanism which can engender 
	superfluidity even for \emph{arbitrarily small} $L$ and $\mu$, as long as $\mu\ne 0$, is a logical possibility. 
	 The condition under which this possibility may be realized  will be carefully examined later. 
	The estimated nonperturbative gap of quarks is of order 
	$\Delta\sim L^{-1}\ee^{-1/(g^4 \mu L)}$.%
	\footnote{It is intriguing to draw a comparison between dense QCD, 
	the NJL model, and our case. In dense QCD the gap is $\sim \ee^{-1/g}$ due to 
	unscreened magnetic gluons \cite{Son:1998uk}. In the NJL-type  
	model with a point-like four-fermion interaction the gap is generally $\sim \ee^{-1/g^2}$ 
	\cite{Buballa:2003qv}. The dependence $\ee^{-1/g^4}$ found in this work differs from 
	either of them.}
\end{itemize}
Now, if the adiabatic continuity of center symmetry at $\mu=0$  
from small to large $S^1$ (conjectured in 
Refs.~\cite{Unsal:2007vu,*Unsal:2007jx,Shifman:2008ja,Argyres:2012ka} 
and tested numerically in Refs.~\cite{Cossu:2009sq,Cossu:2013ora}) holds, 
then it follows that a \emph{double crossover} should emerge on the phase diagram 
(Fig.~\ref{fg:pd}),   
uniting two weakly-coupled BCS superfluids: one at small $S^1$ and 
the other at large $S^1$ and $\mu\gg\Lambda_{\QCD}$. It is highly nontrivial 
that such a continuity may exist, because the two phases have distinct physical 
mechanisms for quark pairing: two-charged-boson exchange interaction on one hand 
(at small $S^1$; see Fig.~\ref{fg:box_dgm} below) 
and the one-gluon-exchange interaction on the other hand. 

%%%%%%%%%%%%%%%%%%%%%%%%%%%%%%%%%%%
%%%%%%%%%%%%%%%%%%%%%%%%%%%%%%%%%%%

This paper is organized as follows. In Sec.~\ref{sc:hoso} we examine 
the perturbative holonomy potential and reveal a complex phase diagram at 
$\mu\ne 0$ and nonzero masses for $\Nc=2$ and $3$. The quark number 
density is also computed. In Sec.~\ref{sc:ssb} we analyze how the monopole-binding 
interaction due to fermion-zero-mode exchange is modified at $\mu\ne 0$. 
It is shown for $\Nc=2$ that the chemical potential renders the inter-monopole potential 
strongly anisotropic, favoring a temporal separation. A fermionic  
low-energy effective theory is also examined and the conditions under which  
spontaneous $\U(1)_B$ breaking occurs at an arbitrarily small chemical potential are derived. 
The fermion gap $\Delta$ is shown to be proportional to  $e^{-1/(g^4 \mu L)}$,  
which is smaller than any finite order 
of the semiclassical expansion in powers of $\ee^{-S_{\rm m}}$ where $S_{\rm m}=8\pi^2/(g^2\Nc)=4\pi^2/g^2$. 
The effective Lagrangian of the Nambu--Goldstone mode associated with 
the $\U(1)_B$ breaking is also derived.  
We conclude in Sec.~\ref{sc:con}. Technical details on 
the derivation of the perturbative potential and properties of a free fermion 
propagator at $\mu\ne 0$ are summarized in Appendix 
\ref{ap:oneloopdet} and \ref{ap:propagator}, respectively.

\section{\label{sc:hoso}Hosotani mechanism with chemical potential}

We consider $\SU(N_c)$ gauge theory on $\RR^3\times S^1$ 
with $\NfD$ flavors of 4-component Dirac fermions 
$\Psi^{}_f = \Psi^{\rmA}_f t^{\rmA}$ in the adjoint representation, obeying PBC on $S^1$. 
We will use $f,g,\dots$ to label flavors $\{1,2,\dots,\NfD\}$ and 
$\rmA, \rmB, \dots$ to label the adjoint colors $\{1,2,\dots,\Nc^2-1\}$. 
(The fundamental colors $\{1,2,\dots,\Nc\}$ will be labeled by $\rma, \rmb, \dots$.) 
The generators are normalized as $\tr(t^\rmA t^\rmB)=\delta^{\rmA \rmB}/2$. 
The circumference of $S^1$ is denoted by $L$. 
The asymptotic freedom requires $\NfD< 2.75$ and this is hereafter assumed 
unless stated otherwise. The Lagrangian reads
\ba
	\mathcal{L} = \tr \ckakko{ 
		\frac{1}{2g^2}F_{\alpha\beta}^2
		+ 2 \bar{\Psi}_f[ \slashed{D}(\mu) + m ]\Psi_f
	},
\ea 
where $F_{\alpha\beta}=F^\rmA_{\alpha\beta}t^\rmA$ and 
$\slashed{D}(\mu)=\slashed{D}-\mu \gamma_4$. 
In this paper we will always work in Euclidean spacetime, 
using $\alpha,\beta,\dots$ to denote the entire four directions  
$\{1,2,3,4\}$ while reserving $i,j,\dots$ for non-compact directions $\{1,2,4\}$ only.  
The compactified spatial direction is $x^3$ in our setting. 

For $\NfD\geq 1$ and $\mu=0$ the gauge symmetry is known to be dynamically broken to 
the maximal torus $\U(1)^{N_c-1}$ 
by quantum effects of periodic fermions if both $L\Lambda_{\QCD}$ and 
$Lm$ are sufficiently small \cite{Hosotani:1988bm,Unsal:2010qh, Unsal:2007vu, Unsal:2007jx}. 
In this section we examine how the chemical potential term $\mu \Psi^\dagger \Psi$ influences 
gauge symmetry breaking at small $S^1$, putting aside the issue of 
photon mass gap and fermion bilinear condensation. This treatment is justified because 
the latter effects are nonperturbatively small while 
gauge symmetry is broken at one loop. 

\subsection{Perturbative potential for the Wilson line}

We consider a uniform background gauge field that is zero in $\RR^3$ directions and 
nonvanishing along $S^1$. The holonomy is given by 
\ba
	\Omega \equiv \mathcal{P}\exp\mkakko{i\oint \dd x^3 A_3}. 
\ea 
We employ the gauge in which $\Omega$ is diagonal,
\ba
	A_3= \mathrm{diag}(a_1,a_2,\dots,a^{}_{\Nc})\,, 
	\quad \sum_{\rma=1}^{\Nc} a_{\rma} =0 
	\label{eq:A3}
\ea
and define dimensionless variables for later convenience:
\ba
	\muu \equiv L\mu,~~\mm\equiv Lm~~~\text{and}~~~
	\aa_\rma \equiv L a_\rma \,.
\ea
The object to be calculated is the one-loop effective potential 
for the holonomy at $\mu \neq 0$. It consists of several pieces,
\ba
	V(\Omega;\mu) & = V_{\rm YM}(\Omega) + 
	\NfD \kkakko{V_{\rm F}(\Omega) + \delta V_{\rm F}(\Omega;\mu)}\,.
	\label{eq:totalV}
\ea
The first term is the contribution of the gauge fields and ghosts \cite{Gross:1980br} 
\ba
	V_{\rm YM}(\Omega) & =\frac{1}{24\pi^2L^4} 
	\sum_{\rma, \rmb=1}^{\Nc} F_1(\aa_{\rma} - \aa_{\rmb}) 
	\label{eq:Veff3}
\ea
with $F_1(\aa) \equiv [\aa]^2 (2\pi-[\aa])^2$ 
where $0\leq [\aa]\leq 2\pi$ is equal to $\aa$ mod $2\pi$\,.  
$V_{\rm F}(\Omega)$ is the potential induced by a single adjoint 
Dirac fermion with PBC at $\mu=0$ \cite{Meisinger:2001fi,Unsal:2010qh},
\ba
	V_{\rm F}(\Omega) & = \frac{2 \mm^2}{\pi^2L^4} 
	\sum_{n=1}^\infty \frac{1}{n^2}|\tr \Omega^n|^2   K_2(n\mm) 
	\notag
	\\
	& = \frac{2}{\pi^2L^4}\sum_{\rma,\rmb=1}^{\Nc} 
	F_2(\mm, \aa_{\rma} - \aa_{\rmb}) 
	\label{eq:Veff2}
\ea
with
\ba
	F_2(\mm,\aa) \equiv \mm^2 \sum_{n=1}^{\infty}\frac{1}{n^2}K_2(n\mm)\cos(n\aa) \,.
\ea

What remains is to compute $\delta V_{\rm F}(\Omega;\mu)$, which controls the 
$\mu$-dependence of the effective potential. In terms of the quark determinant 
\ba
	\Gamma_q(\Omega;\mu) & \equiv \log \det[\DD(\mu)+m]\,,
\ea
we have 
\ba
	\delta V_{\rm F}(\Omega;\mu) = - \frac{1}{V_{\RR^3}L}
	\ckakko{ \Gamma_q (\Omega;\mu) - \Gamma_q (\Omega;0) } 
\ea
where $V_{\RR^3}$ formally denotes the infinite 
volume of the spacetime in $x^{1,2,4}$ directions. Technical details of 
the evaluation of $\Gamma_q$ is given in Appendix \ref{ap:oneloopdet}. 
The result is
\ba
	\delta V_{\rm F}(\Omega;\mu) = 
	- \frac{1}{2\pi L^4} \bigg\{
		\sum_{\rma,\rmb=1}^{\Nc}F_3(\muu,\mm,\aa_{\rma} - \aa_{\rmb}) 
		- F_3(\muu,\mm,0)
	\bigg\}
	\label{eq:Veff1}
\ea
with 
\ba
	F_3(\muu,\mm,\aa) & \equiv \sum_{n=-\infty}^{\infty}
	\theta\big( \muu^2 - \mm^2 - \left(\aa+2n\pi \right)^2 \big)
	\notag
	\\
	& \quad \times 
	\bigg[
		\frac{1}{3}\muu^3 -\muu \big\{( \aa + 2n\pi )^2+\mm^2 \big\}  
	\notag
	\\
	& \quad + \frac{2}{3}\big\{( \aa + 2n\pi )^2+\mm^2 \big\}^{3/2}
	\bigg] \,,
	\label{eq:F333}
\ea
where $\theta(x)$ is the Heaviside step function. 
It is immediately clear from \eqref{eq:F333} that 
$\delta V_{\rm F}(\Omega;\mu)$ vanishes when $\mu<m$. 
This is expected because $\mu$ below the energy gap (mass of adjoint fermion) has no physical effect 
at zero temperature. 

Collecting \eqref{eq:totalV}, \eqref{eq:Veff3}, \eqref{eq:Veff2} and \eqref{eq:Veff1} 
we obtain the one-loop effective potential. Next we proceed to numerical 
analysis of the phase diagram for $N_c=2$ and $3$ based on minimization of $V(\Omega;\mu)$.

%HERE>......HERE>......HERE>......HERE>......HERE>......

\subsection{\boldmath $N_c=2$}

For $\Nc=2$ the holonomy can be parametrized as $\Omega=\diag(\ee^{i\aa},\ee^{-i\aa})$. 
In numerical minimization of the potential $V(\Omega;\mu)$ we encounter 
the following two phases. 
\begin{center}
	\begin{tabular}{|c||c|c|}
		\hline 
		minimum & ~$\ZZ_2$ center~ & gauge symmetry
		\\\hline 
		$\aa=0,\pi$ & broken & $\SU(2)$
		\\
		$\aa=\pi/2$ & unbroken & $\U(1)$
		\\\hline 
	\end{tabular}
\end{center}
Let us summarize the bosonic spectrum in each phase.
\begin{itemize}
	\item 
	In the first phase with broken center symmetry, 
	the $A_3$ component of gluons acquires a mass 
	of $\calO(g/L)$ (called the electric screening mass 
	in the case of thermal $S^1$) while the other components $A_{1,2,4}$ 
	are perturbatively massless and  
	interact strongly with periodic fermions. In the limit $L\to 0$ with fixed $\mu$ 
	the system would reduce to three-dimensional adjoint QCD at finite $\mu$.
	\item 
	The second phase entails Abelianization of the gauge group due to the Hosotani mechanism;  
	similarly to $W$ bosons in the electroweak theory, 
	$A_{1,2,4}^{\rmA=1,2}$ acquire a mass $m^{}_{\rm W}=\pi/L$ by eating $A_3^{\rmA=1,2}$;  
	the ``Higgs mode'' $A_3^{\rmA=3}$ gains a mass $m^{}_{\rm H}\sim g/L$;   
	The photons $A_{1,2,4}^{\rmA=3}$ are massless to all orders in perturbation theory. 
	\end{itemize}

\begin{figure}[tb]
	\centering
	\includegraphics[width=.47\textwidth]{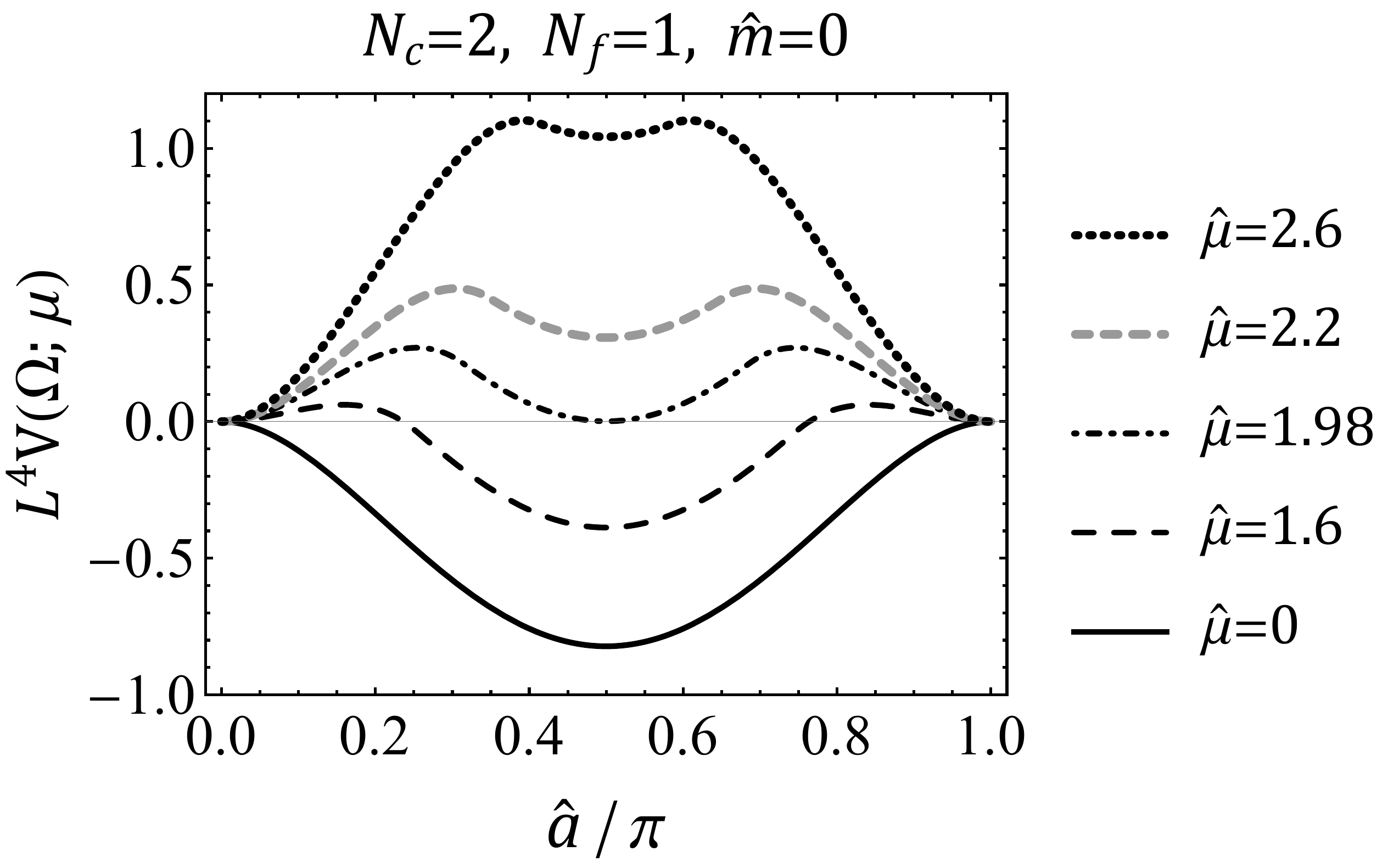}
	\vspace{-2mm}
	\caption{\label{fg:potential_Nc2_Nf1_m0}%
	Effective potential for $N_c=2$ and $\NfD=1$ 
	in the chiral limit. In this figure $V$ is normalized to $0$ at $\aa=0$.
	}
\end{figure}

In Fig.~\ref{fg:potential_Nc2_Nf1_m0} 
the holonomy potential for $\NfD=1$ in the chiral limit is plotted. 
As $\muu$ increases, the center-symmetric phase is turned into a center-broken phase via a 
first-order transition. Namely $\mu$ counteracts the Hosotani mechanism in this parameter range.% 
\footnote{This is in contrast with the case of an external magnetic field \cite{Anber:2013tra}, 
which \emph{enlarges} the center-symmetric phase.} 
%%%%%%%%%%%%%%%%%%%%%%%%%%%%%%%%%%%
%%%%%%%%%%%%%%%%%%%%%%%%%%%%%%%%%%%
To explain why, let us look at the quark number density  
\ba
	n_q & \equiv - \NfD \frac{\der}{\der\mu}\delta V_{\rm F}(\Omega;\mu)
	\notag \\
	& = \frac{\NfD}{2\pi L^3} \big\{
		2 F'_3(\muu, \mm, 2\aa) + F'_3(\muu,\mm,0) 
	\big\}
	\label{eq:nq}
\ea
with
\ba
	F_3'(\muu,\mm,\aa) & = \sum_{n=-\infty}^{\infty}
	\{ \muu^2 - \mm^2 - \left(\aa+2n\pi \right)^2 \}
	\notag
	\\
	& \quad \times  
	\theta\big( \muu^2 - \mm^2 - \left(\aa+2n\pi \right)^2 \big)\,.
	\label{eq:F3p}
\ea
The first term in \eqref{eq:nq} represents the density of $\Psi^{\rmA=1,2}$, 
which are charged under the unbroken $\U(1)$ gauge symmetry. The second term in \eqref{eq:nq} 
represents the density of neutral fermions $\Psi^{\rmA=3}$; since it does not depend on 
the holonomy it can be ignored for the moment. 

\begin{figure}[tb]
	\centering
	\includegraphics[width=.4\textwidth]{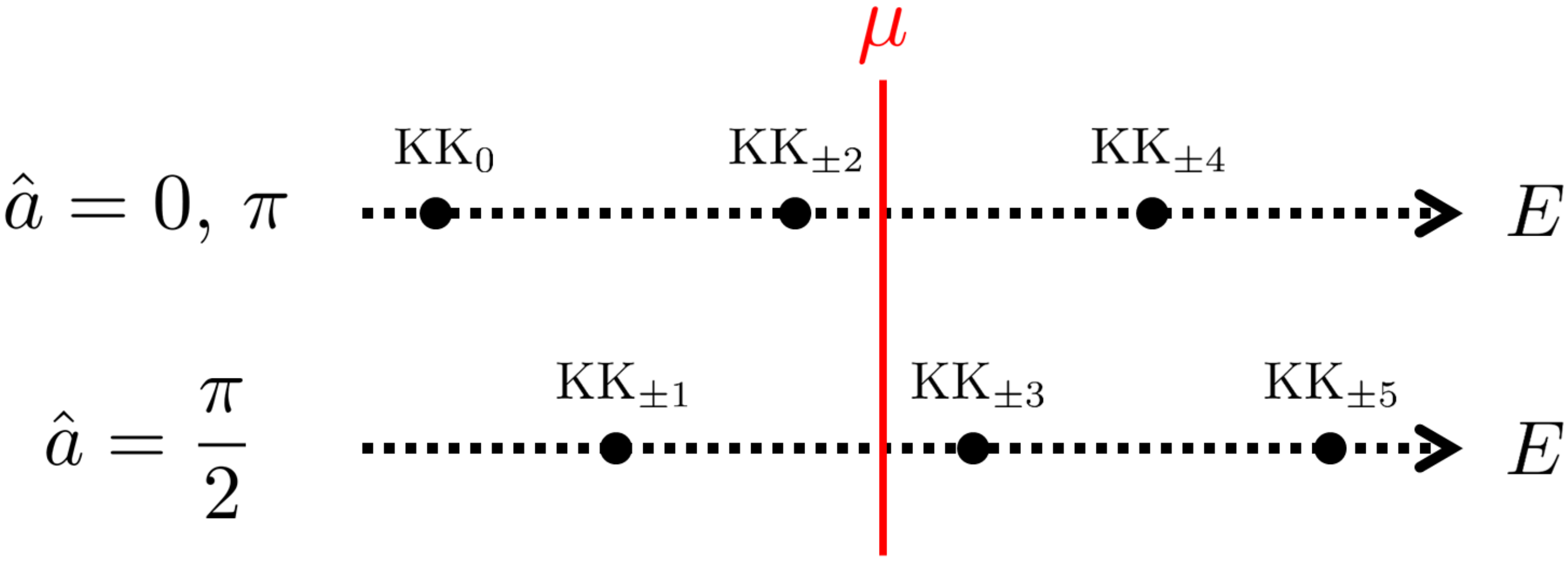}
	\caption{\label{fg:KK}%
	The structure of $\KK_\ell$ energy levels for the center-broken ($\aa=0,\pi$) 
	and center-symmetric ($\aa=\pi/2$) phases. In this example, the Fermi sea 
	of charged fermions consists of 
	$\KK_{0}\cup\KK_{\pm 2}$ or $\KK_{\pm 1}$, depending on the holonomy.}
\end{figure}

As \eqref{eq:F3p} implies, due to compactification,  
the Fermi sea becomes a union of multiple disks in momentum space, labeled as
\ba
	\KK_\ell \equiv \left\{
		(p_1,p_2,p_3) \,\bigg|\; p_3=\frac{\ell \pi}{L}, \;  
		\sqrt{\bm{p}^2+m^2}\leq \mu
	\right\}
\ea
with $\ell\in\ZZ$. (Precisely speaking, this $p_3$ is the eigenvalue of $i\der_3+A_3$ 
rather than $i\der_3$.) Such a discrete structure of energy levels is 
analogous to the Landau levels in a magnetic field. As shown in Fig.~\ref{fg:KK}, charged fermions occupy either 
$\KK_{\ell=\text{even}}$ or $\KK_{\ell=\text{odd}}$ with 
$|\ell|\leq \sqrt{\muu^2-\mm^2}/\pi$ depending on the holonomy. Namely, 
in the center-symmetric phase $(\aa=\pi/2)$ the holonomy dynamically 
changes the boundary condition of charged fermions along $S^1$ from periodic 
to \emph{anti-periodic}. We note that every time $\mu$ reaches a new $\KK_\ell$ level, 
a second-order phase transition occurs; the quark number susceptibility jumps. 
Indeed there are an infinite number of second-order transitions as a function of 
$\mu$.%
\footnote{This phenomenon, analogous to the de Haas--van Alphen effect 
in solid state physics, has been studied in NJL models with spatial compactification 
\cite{Vshivtsev:1998fg}.} 
%%%%%%%%%%%%%%%%%%%%%%%%%%%%%%%%%%%
%%%%%%%%%%%%%%%%%%%%%%%%%%%%%%%%%%%

Let us examine what happens at small $\muu$. When 
$\mm\leq \muu<\sqrt{\pi^2+\mm^2}$, $\mu$ lies between $\KK_0$ 
and $\KK_{\pm 1}$. Hence $\KK_0$ forms the charged Fermi sea 
in the center-broken phase. On the other hand, no Fermi sea exists yet 
in the center-symmetric phase. This means that the pressure in the 
center-broken phase is higher, by the amount of the degenerate Fermi sea, 
implying that center symmetry breaking is favored by $\delta V_{\rm F}(\Omega;\mu)$ 
for $\muu$ in this range. This explains the first-order transition in 
Fig.~\ref{fg:potential_Nc2_Nf1_m0}.

\begin{figure}[tb]
	\centering
	\includegraphics[width=.35\textwidth]{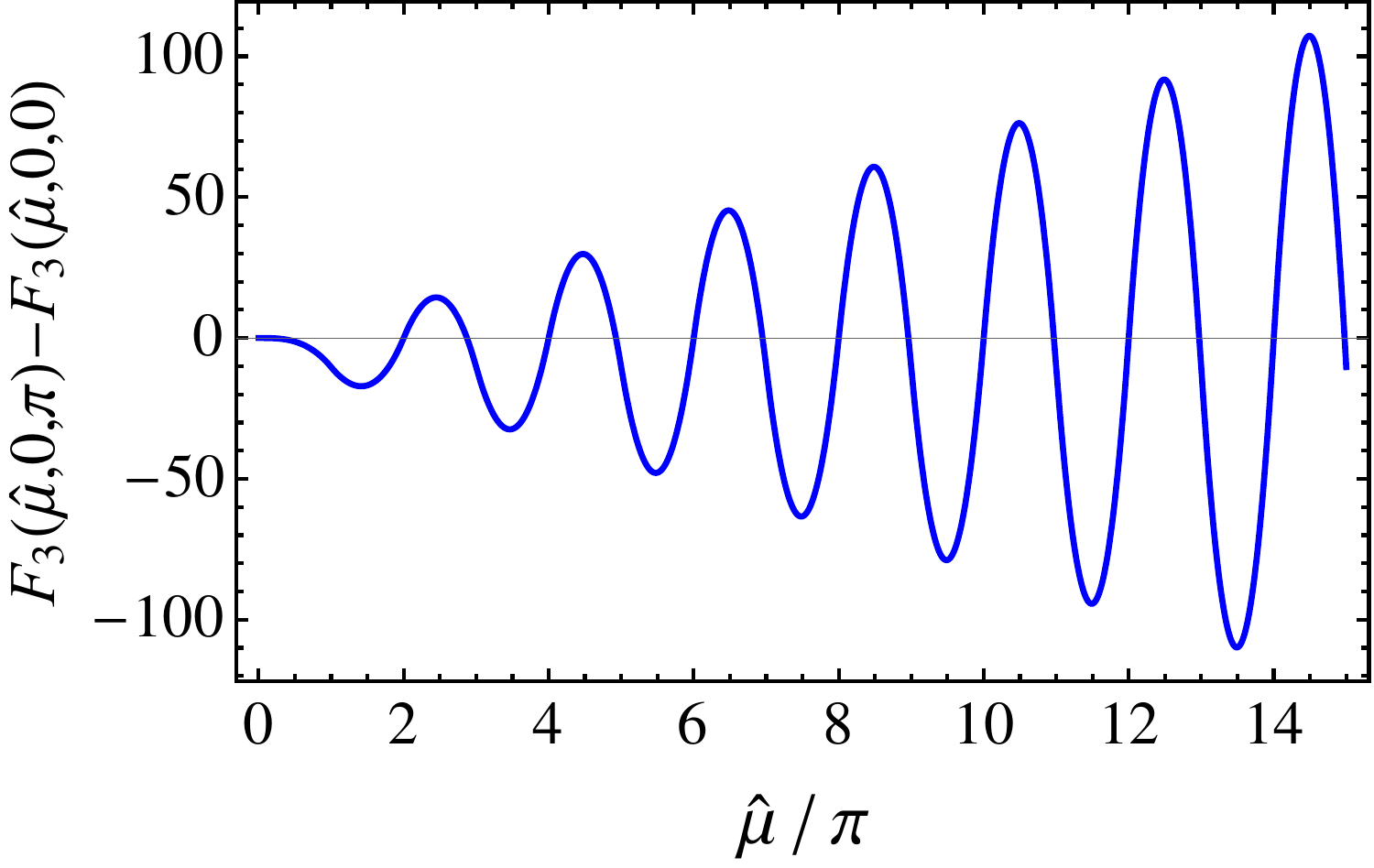}
	\vspace{-3mm}
	\caption{\label{fg:F3_oscillation}%
	Difference of the free energy with a center-symmetric vs.~center-broken 
	holonomy in the massless limit [with $F_3$ defined in \eqref{eq:F333}].}
\end{figure}

\begin{figure}[tb]
	\centering
	\includegraphics[width=.4\textwidth]{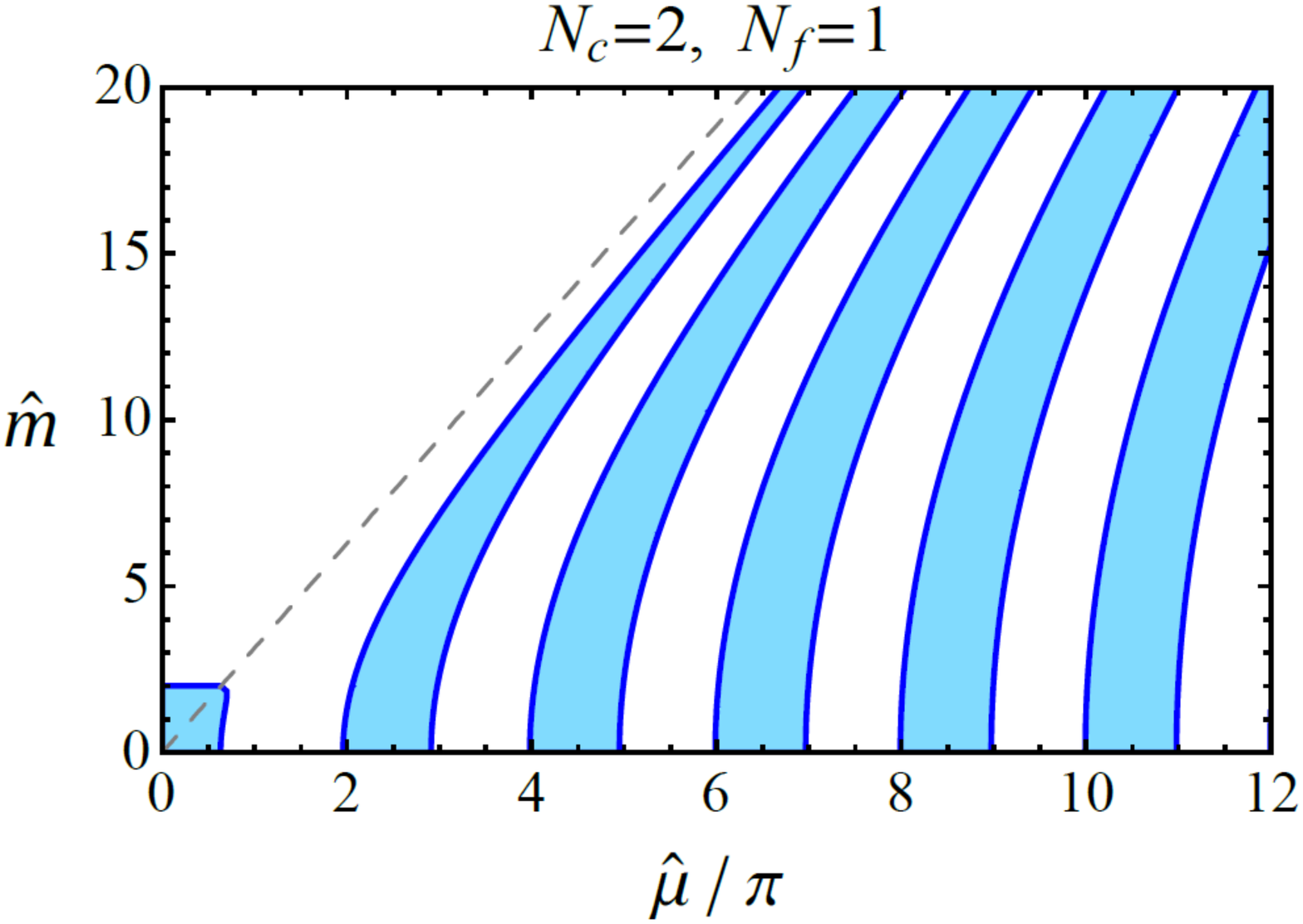}
	\vspace{-3mm}
	\caption{\label{fg:pd_Nc2_Nf1}%
	Phase diagram of center symmetry for $\Nc=2$ and $\NfD=1$. 
	The shaded (empty) region is a center-symmetric (center-broken) 
	phase, respectively. They are separated by 
	first-order transitions. 	The dashed line represents $\mm=\muu$. }
\end{figure}

Interestingly, this is not the whole story. As $\muu$ increases further, 
many $\KK_\ell$ join the Fermi sea and the competition between 
$\KK_{\ell=\text{even}}$ and $\KK_{\ell=\text{odd}}$ becomes quite subtle. 
In Fig.~\ref{fg:F3_oscillation} we plot the difference of $F_3$, \eqref{eq:F333}, 
in the two vacua at $\mm=0$.  
%%%%%%%%%%%%%%%%%%%%%%%%%%%%%%%%%%%
%%%%%%%%%%%%%%%%%%%%%%%%%%%%%%%%%%%
It shows a growing oscillation in $\muu$ with period $\approx 2\pi$.  
At small $\muu$ the oscillation starts with a negative value, implying that 
the center-broken vacuum is favored by medium. As $\muu$ increases, however, 
the zero is crossed infinitely many times and consequently it leads to an 
endless alternation of the two vacua. In Fig.~\ref{fg:pd_Nc2_Nf1} we show 
the phase diagram of center symmetry, showing the presence of infinitely 
many center-symmetric phases. 
%%%%%%%%%%%%%%%%%%%%%%%%%%%%%%%%%%%
%%%%%%%%%%%%%%%%%%%%%%%%%%%%%%%%%%%
One can tune $\muu$ to bring the system into a center-symmetric phase, 
however large the mass $\mm$. 

It should be noted that  the center-symmetric phases at $\muu>1$ 
are different from the well-studied center-symmetric phase at $\mm\sim \muu\sim 0$ 
in that there is a Fermi sea of charged fermions $\Psi^{\rmA=1,2}$ 
in the former but not in the latter. 

\begin{figure}[tb]
	\centering
	\includegraphics[width=.41\textwidth]{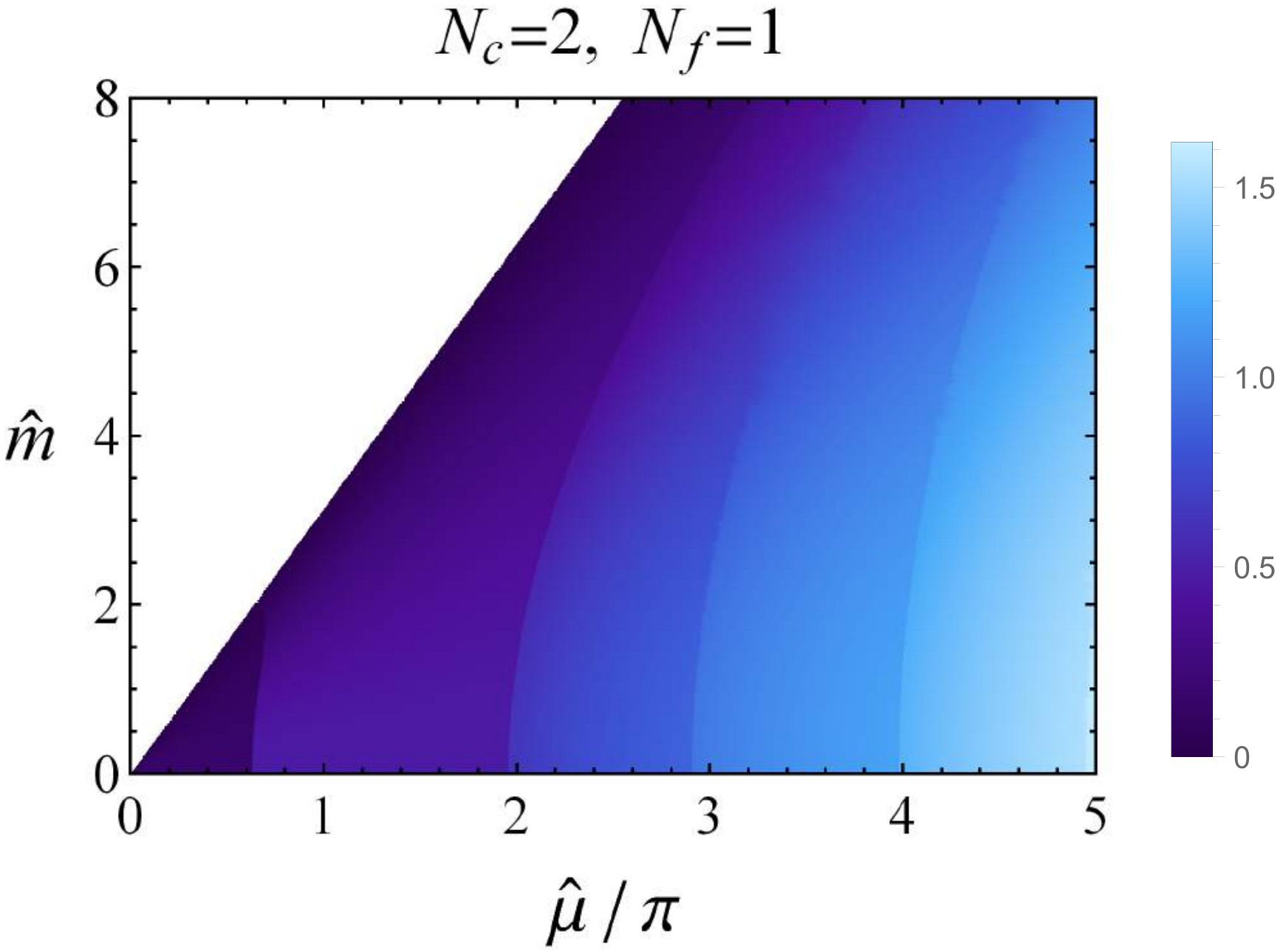}
	\vspace{-3mm}
	\caption{\label{fg:quark_num_density_Nc2}%
	The rescaled quark number density $Ln_q/\mu^2$ for $\Nc=2$ and $\NfD=1$. 
	The $\mm>\muu$ region where the density is strictly zero is left empty.  
	}
\end{figure}

The stripe structure of the phase diagram in Fig.~\ref{fg:pd_Nc2_Nf1} 
is reflected in the $\mu$-dependence of other observables as well. 
Figure~\ref{fg:quark_num_density_Nc2} plots the quark density $n_q$   
(divided by $\mu^2$ to make the stratified structure clearer). 
The lines of center-symmetry-changing first-order transitions are clearly visible. 
In addition, $m^{}_{\rm H}$ (the mass of $A_3^{\rmA=3}$) oscillates with $\muu$, but 
it stays at $\calO(g/L)$ and never vanishes because the transitions are first order. 
%%%%%%%%%%%%%%%%%%%%%%%%%%%%%%%%%%%
%%%%%%%%%%%%%%%%%%%%%%%%%%%%%%%%%%%

\subsection{\boldmath $N_c=3$}

For $\Nc=3$ the holonomy can be parametrized as 
$\Omega=\diag\big(\ee^{i\aa_1},\ee^{i\aa_2},\ee^{-i(\aa_1+\aa_2)}\big)$. 
In numerical minimization of the potential $V(\Omega;\mu)$ with respect to $\aa_1$ and $\aa_2$ 
we found three phases below. 
\begin{center}
	\scalebox{0.9}{$\displaystyle 
	\begin{tabular}{|c||c|c|}
		\hline 
		$ (\aa_1,\aa_2)$ at the minimum & $\ZZ_3$ center & gauge symmetry
		\\\hline 
		$(0,0)$, $\pm\mkakko{\frac{2\pi}{3},\frac{2\pi}{3}}$ & broken & $\SU(3)$ 
		\!\!{\Large \strut}
		\\
		$\pm\mkakko{\frac{\pi}{3},\frac{\pi}{3}}, 
		\pm\mkakko{\frac{\pi}{3},-\frac{2\pi}{3}}, 
		\pm\mkakko{-\frac{2\pi}{3},\frac{\pi}{3}}$ & broken & $\SU(2)\times\U(1)$ 
		\!\!{\Large \strut}
		\\
		$\mkakko{0,\pm\frac{2\pi}{3}}, \mkakko{\pm\frac{2\pi}{3},\mp\frac{2\pi}{3}}, 
		\mkakko{\pm\frac{2\pi}{3},0}$   & unbroken & $\U(1)\times\U(1)$ 
		\!\!{\Large \strut}
		\\\hline 
	\end{tabular}
	$}
\end{center}
The phase in the second row, often called the \emph{split} (or \emph{skewed}) phase, was 
found in Ref.~\cite{Myers:2007vc} in a deformed $\SU(3)$ Yang-Mills theory and 
has been studied in a variety of setups \cite{Myers:2009df,Cossu:2009sq,Cossu:2013ora,Misumi:2014raa}. 
The other two phases are natural generalizations of those for $\Nc=2$. 

\begin{figure}[tb]
	\centering
	\includegraphics[width=.49\columnwidth]{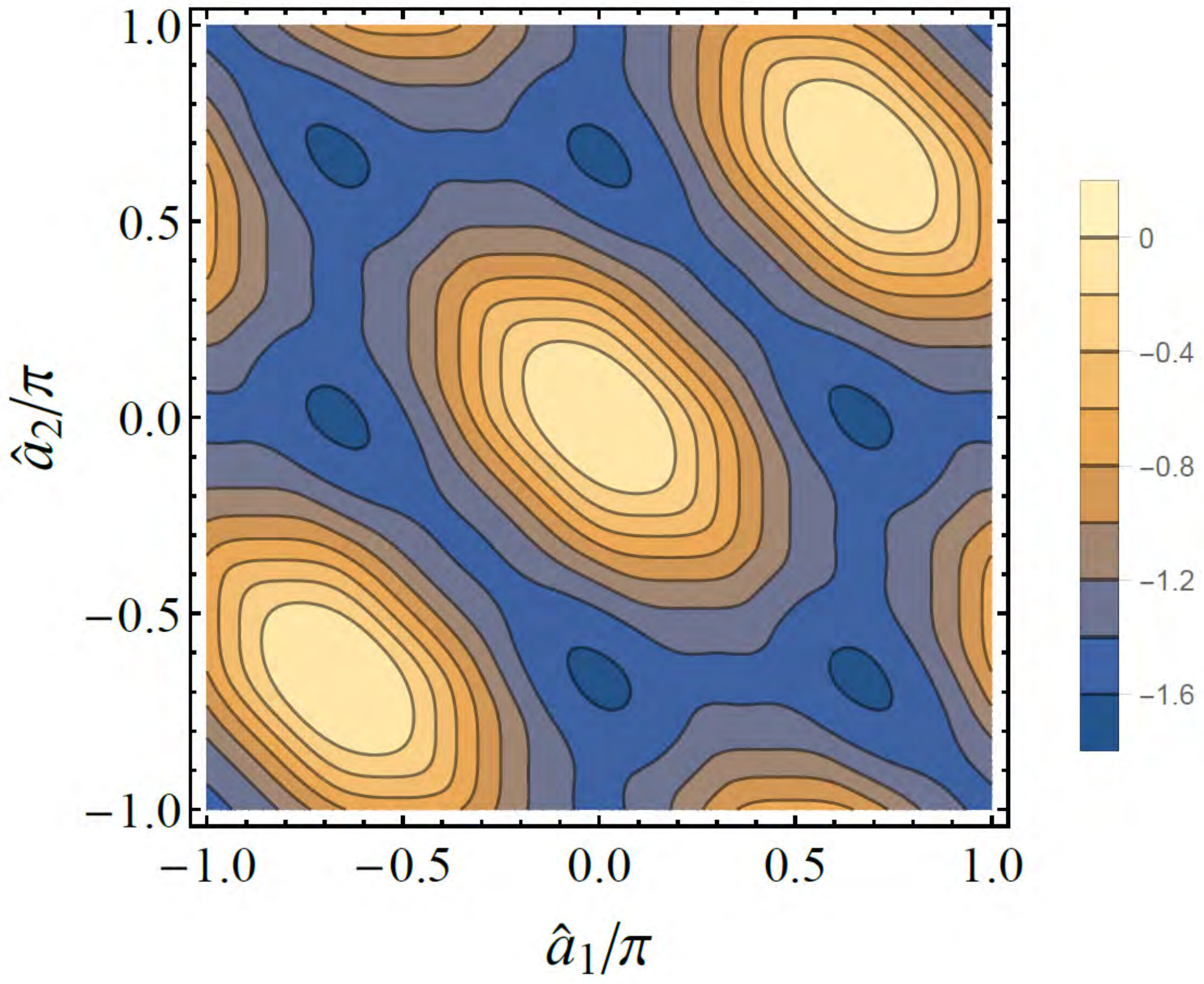}
	\includegraphics[width=.49\columnwidth]{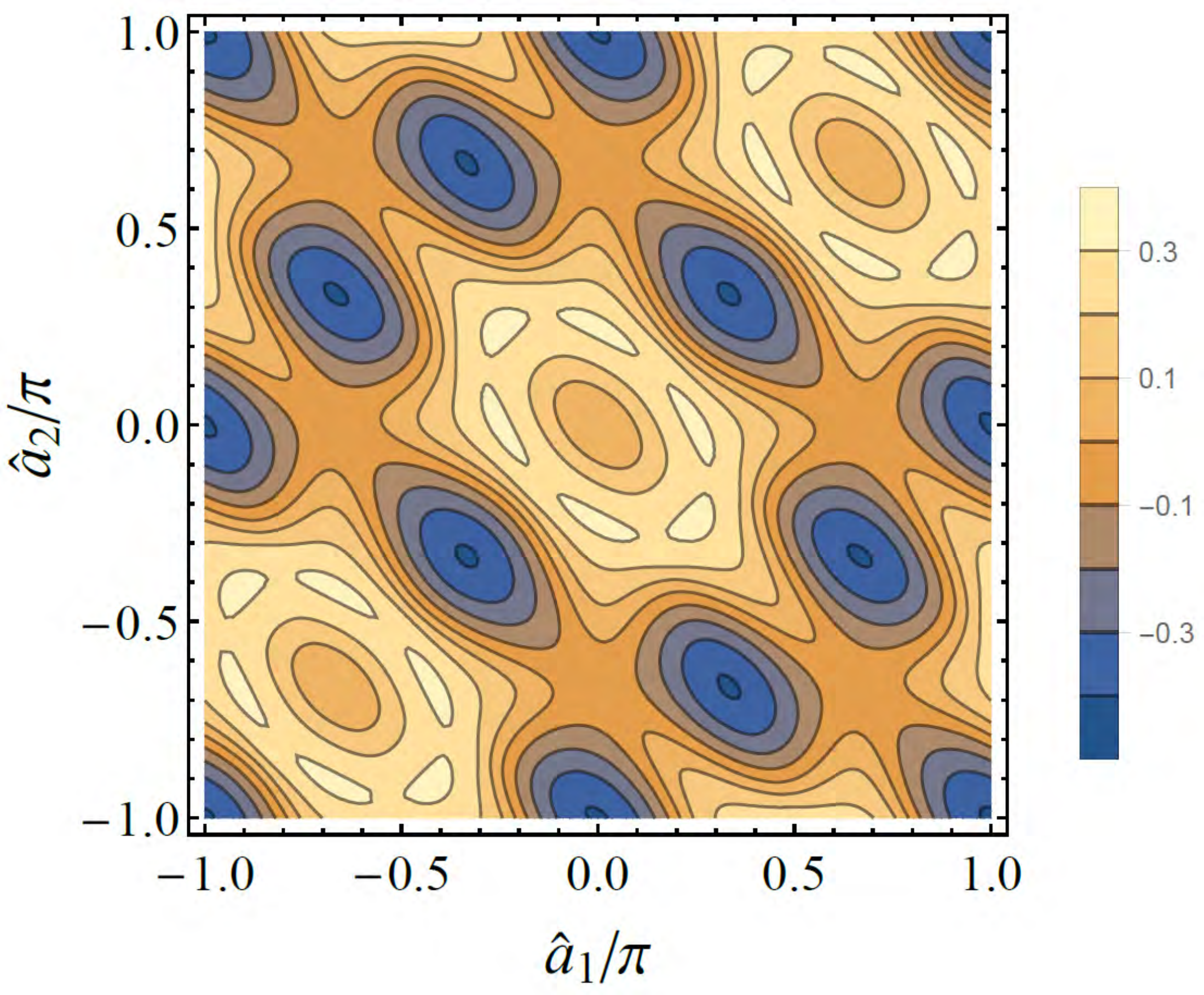}
	\put(-198,102){$\muu=1.0$}
	\put(-75,102){$\muu=1.8$}
	\vspace{5pt}\\
	\includegraphics[width=.49\columnwidth]{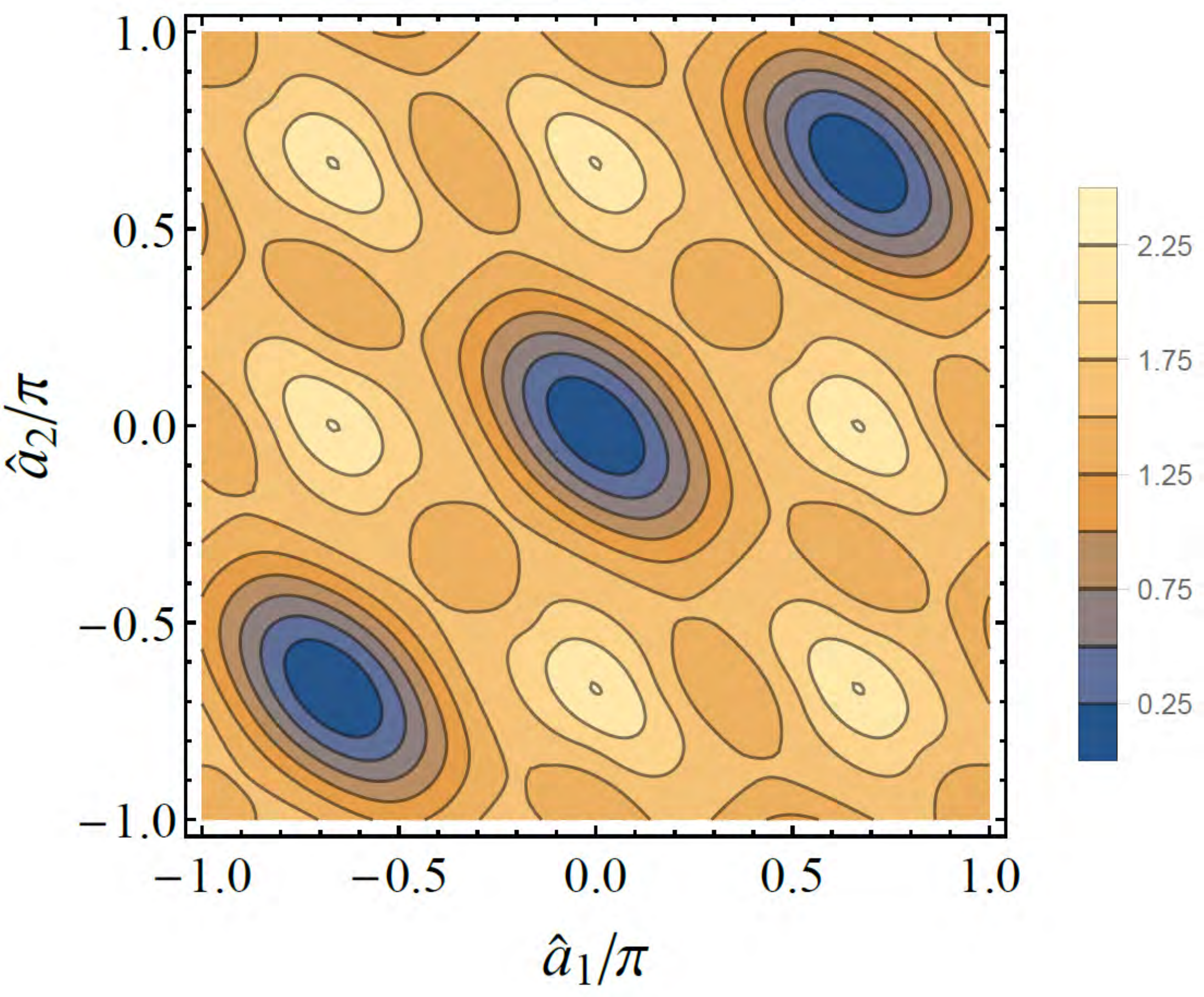}
	\put(-75,103){$\muu=2.4$}
	\vspace{-2mm}
	\caption{\label{fg:potential_Nc3}%
	Effective potential $L^4V(\Omega;\mu)$ 
	for $\Nc=3$ and $\NfD=1$ in the chiral limit, normalized to $0$ at $\aa_1=\aa_2=0$. 
	Each $\muu$ corresponds to the three different phases (see the main text and 
	Fig.~\ref{fg:pd_Nc3_Nf1}). First-order phase transitions occur at 
	$\muu=1.42$ and $\muu=1.98$. 
	}
\end{figure}

As for $\Nc=2$, we again find that the addition of $\muu$ destabilizes the 
center-symmetric vacuum realized at $\muu=0$.  Figure~\ref{fg:potential_Nc3} 
displays the evolution of the holonomy potential for increasing $\muu$.  We observe that 
each phase in the above table is realized successively (from bottom to top) for increasing $\muu$. 
An explanation in terms of $\KK_\ell$ levels becomes rather cumbersome and 
is not as useful as for $\SU(2)$. 

%%%%%%%%%%%%%%%%%%%%%%%%%%%%%%%%%%%
%%%%%%%%%%%%%%%%%%%%%%%%%%%%%%%%%%%

\begin{figure}[tb]
	\centering
	\includegraphics[width=.4\textwidth]{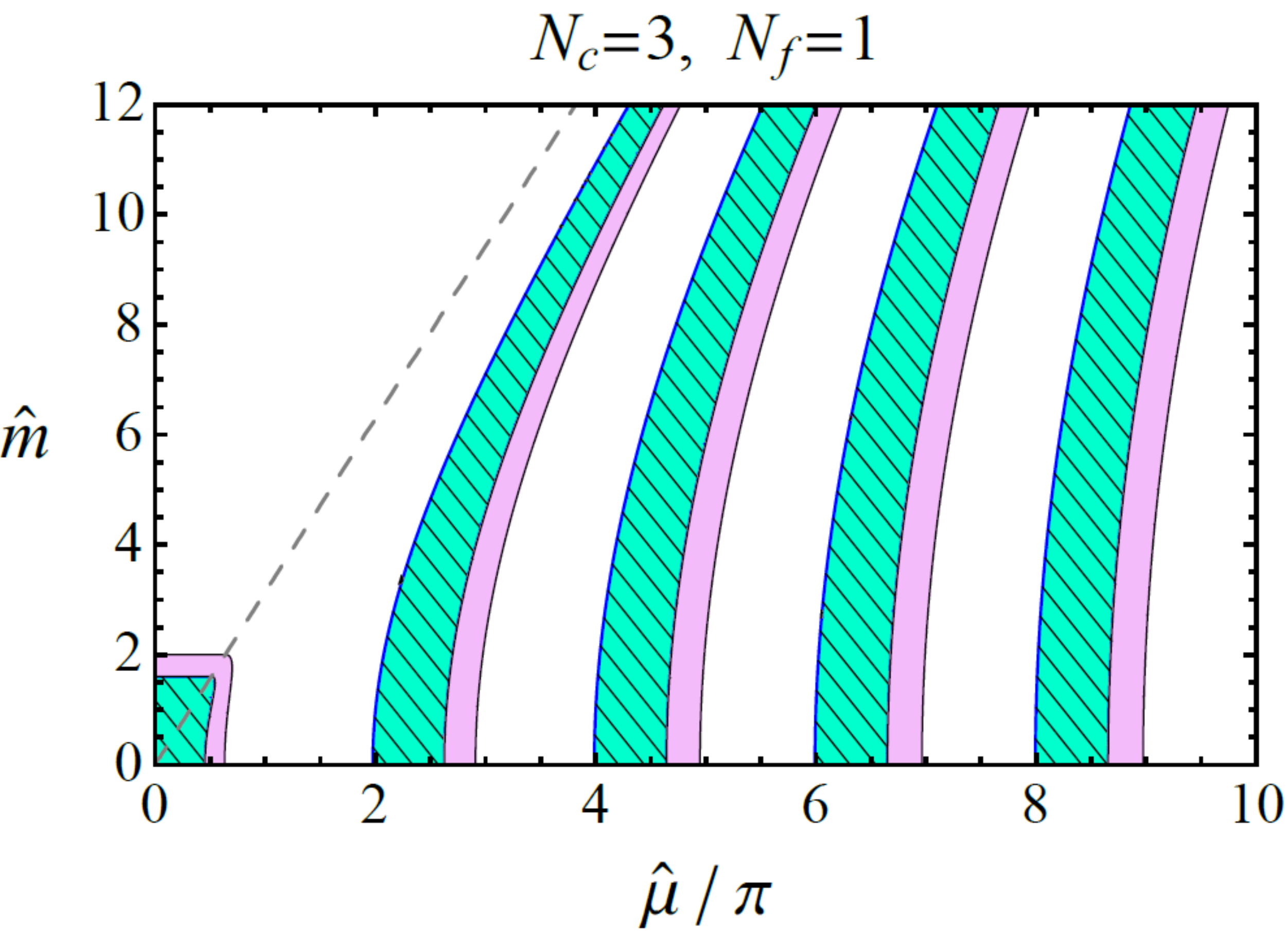}
	\vspace{-3mm}
	\caption{\label{fg:pd_Nc3_Nf1}%
	Phase diagram for $\Nc=3$ and $\NfD=1$. 
	The hatched region is the $\U(1)\times\U(1)$ center-symmetric phase, the shaded region 
	is the $\SU(2)\times \U(1)$ split phase and the empty region is the $\SU(3)$ center-broken phase. 
	The phase boundaries are first order. The dashed line represents $\mm=\muu$. 
	}
\end{figure}

\begin{figure}[tb]
	\centering
	\includegraphics[width=.41\textwidth]{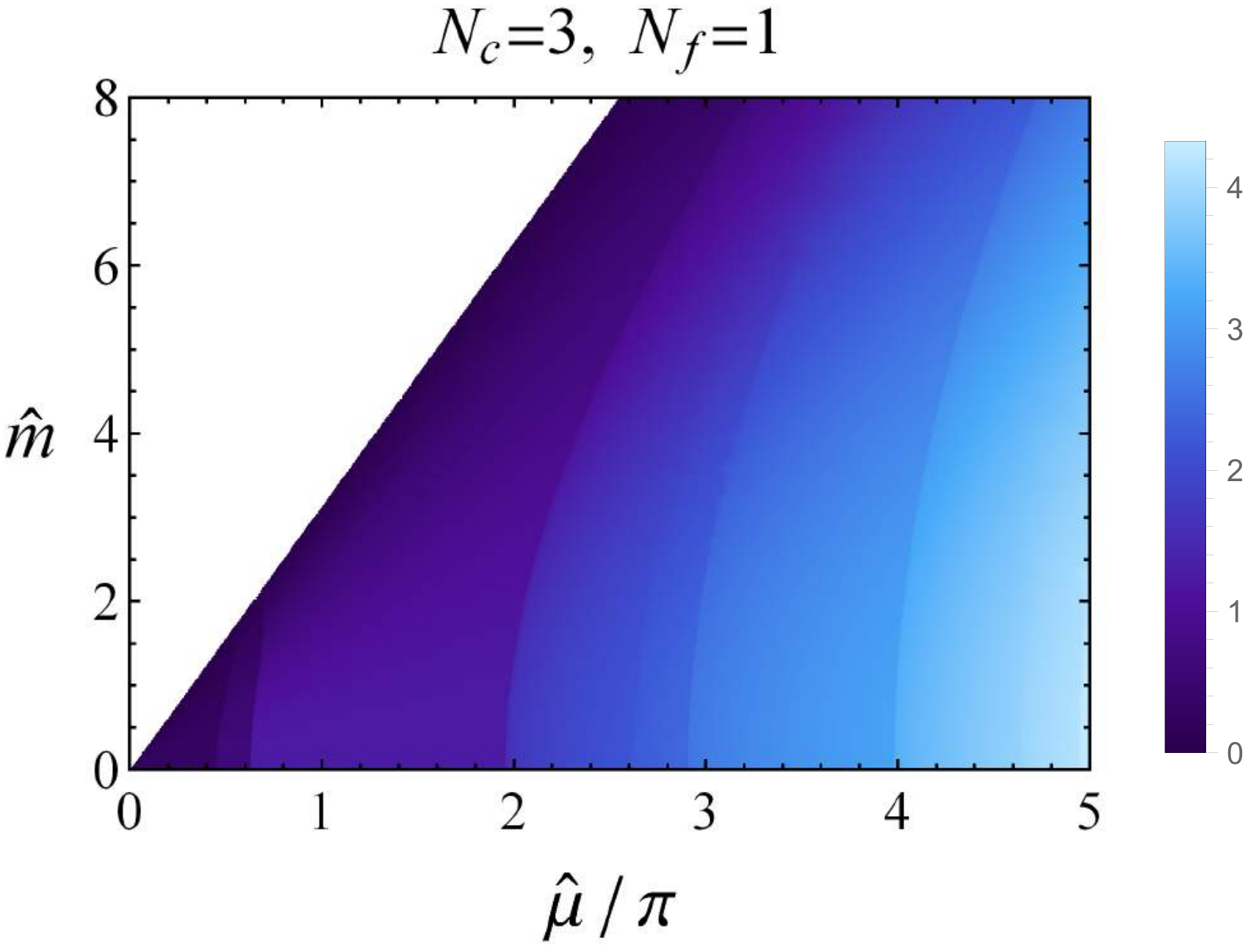}
	\vspace{-3mm}
	\caption{\label{fg:quark_num_density_Nc3}%
	The rescaled quark number density $Ln_q/\mu^2$ for $\Nc=3$ and $\NfD=1$. 
	The $\mm>\muu$ region where the density is strictly zero is left empty.  
	}
\end{figure}

When $\muu$ is further increased, the three phases periodically alternate 
and there are an infinite number of first-order transitions. The phase diagram 
for $\NfD=1$ is displayed in Fig.~\ref{fg:pd_Nc3_Nf1}. Again we find that 
a center-symmetric phase can be realized for an \emph{arbitrarily large} fermion mass 
by tuning $\mu$ judiciously. 
The global features of Fig.~\ref{fg:pd_Nc3_Nf1} 
are similar to Fig.~\ref{fg:pd_Nc2_Nf1} except for the presence of the split phase. 
%%%%%%%%%%%%%%%%%%%%%%%%%%%%%%%%%%%
%%%%%%%%%%%%%%%%%%%%%%%%%%%%%%%%%%%
The quark density $n_q/\mu^2$ is plotted in Fig.~\ref{fg:quark_num_density_Nc3}. 
It increases monotonically with $\muu$ and exhibits tiny jumps 
along the first-order transition lines in Fig.~\ref{fg:pd_Nc3_Nf1}. 
%%%%%%%%%%%%%%%%%%%%%%%%%%%%%%%%%%%
%%%%%%%%%%%%%%%%%%%%%%%%%%%%%%%%%%%

It is tempting to conjecture that, as we increase $\Nc$, the phase diagram 
will contain a variety of more exotic phases with partial center 
symmetry breaking, such as those found in 
Refs.~\cite{Myers:2009df,Unsal:2010qh,Bringoltz:2011by}. However, when 
$L$ is made larger, all such center-breaking phases must somehow disappear 
because adjoint QCD on $\RR^4$ is a strongly-coupled confining theory 
(cf.~Fig.~\ref{fg:pd}). 
Although revealing the phase structure at intermediate $L$ is beyond the realm of 
weak-coupling method in this paper, this can be studied in lattice 
simulations in principle as the path-integral measure of 
adjoint QCD is positive semi-definite even at $\mu\ne 0$ \cite{Alford:1998sd}.

%HERE>......HERE>......HERE>......HERE>......HERE>......

\section{\label{sc:ssb}Low-energy effective theory and symmetry breaking}

\subsection{General discussion}

In the previous section we have shown by using the one-loop effective potential 
of holonomy that there are three phases (below we assume $\Nc=2$ for simplicity):
\begin{itemize} 
	\item Phase I: Center-broken phase with $\Omega=\pm \1_2$ 
	\item Phase II: Center-symmetric phase at $\muu>\pi$
	\item Phase III: Center-symmetric phase at $\muu<\pi$ 
	(the small square next to the origin in Fig.~\ref{fg:pd_Nc2_Nf1})
\end{itemize}
Phases II and III are similar in terms of center symmetry, but it is useful to distinguish them 
for later purposes. In the following, we aim to analyze the low-energy dynamics in each phase, with 
a focus on Phase III. Before doing so, it may be useful to remind global 
symmetries of adjoint QCD as it has unique features absent in ordinary QCD. 
The classical massless Lagrangian of this theory with $\NfD$ Dirac flavors  
at $\mu=0$ is invariant under $\U(1)_A\times\SU(2\NfD)$ symmetry due to reality of 
the adjoint representation \cite{Peskin:1980gc,Leutwyler:1992yt}, 
with $\U(1)_B\subset \SU(2\NfD)$ the baryon number symmetry 
acting as $\Psi\to \ee^{i\varphi}\Psi$. Chiral anomaly reduces $\U(1)_A$ down to 
$\ZZ_{4N_c\NfD}$. In $\RR^4$ (i.e., $L\to\infty$), if $\NfD$ is below 
the conformal window, the continuous part of the flavor symmetry will be 
spontaneously broken by chiral condensate $\langle\tr\bar\Psi\Psi\rangle\ne0$ 
as $\SU(2\NfD)\to\SO(2\NfD)$ 
\cite{Peskin:1980gc,Leutwyler:1992yt}. The chemical potential $\mu \neq 0$ spoils  
the symmetry between fermions and antifermions and breaks the flavor symmetry 
explicitly as $\SU(2\NfD)\to \U(1)_B\times \SU(\NfD)_L\times\SU(\NfD)_R$ 
for $\NfD\geq 2$ and $\SU(2\NfD)\to \U(1)_B$ for $\NfD=1$ \cite{Kogut:2000ek}. 
From here on, we will focus our attention on the minimal $\NfD=1$ case, so that 
the non-anomalous global symmetry that can be spontaneously 
broken at $\mu \neq 0$ is $(\ZZ_8)_A \times \U(1)_B$, as noted earlier in \eqref{eq:sym3}. 

Let us begin with Phase I in the list above. This phase, having intact $\SU(2)$ 
gauge symmetry, bears similarity to the deconfined phase of thermal QCD, except 
that fermions obey PBC along $S^1$ here. In the limit $L\to0$ all modes other than 
$\KK_0$ will decouple and we end up with 
adjoint QCD in $\RR^3$ with $\mu$, with a dimensionful coupling%
\footnote{Notice that the scale of $\mu$ is extremely high in this dimensionally 
reduced theory: $\mu/g_3^2=\muu/g^2\sim 1/g^2\gg 1$.} 
\ba
\label{g3}
g_3^2\equiv \frac{g^2}{L}\,.
\ea
This is a strongly coupled theory and not amenable to semiclassical analysis, but 
we can guess the physics qualitatively. Because there is a nonzero density of fermions 
for $\mu>m$ and there is an attractive channel in the color  
interaction mediated by $\SU(2)$ gluons, the BCS mechanism  
will most likely trigger diquark condensation $\langle\tr\Psi^{\rm T}\Psi\rangle$ 
that breaks $\U(1)_B$ spontaneously and generates a pairing gap for fermions. 
While this scenario sounds plausible, a serious study must incorporate the pairing of 
fermions and the  nonperturbative mass gap of gluons simultaneously. 
This is an interesting open problem.  

Next we consider Phase II. At energy scales far below 
$m^{}_{\rm W} = \pi/L$ and $m^{}_{\rm H}=\calO(g/L)$, 
one can adopt a description in terms of compact $\U(1)$ gauge 
theory with fermions $\Psi$. At $\mu=0$, charged fermions 
$\Psi^{\pm}\equiv (\Psi^{\rmA=1}\pm i \Psi^{\rmA=2})/\sqrt{2}$ can also be 
integrated out due to their large ``screening mass'' $\pi/L$ 
\cite{Unsal:2007vu,Unsal:2007jx}. However, at 
$\mu>\sqrt{(\pi/L)^2+m^2}$ (or $\muu>\pi$ in the chiral limit), 
there is a Fermi surface of $\Psi^\pm$ on top of that of neutral $\Psi^{\rmA=3}$, 
so the charged fermions cannot be dropped from the long-distance effective description.%
\footnote{This situation is analogous to center-stabilized QCD with fundamental fermions in 
$\RR^3\times S^1$ \cite{Shifman:2008ja,Poppitz:2013zqa}.} 
The attractive Coulomb interaction between $\Psi^{+}$ and $\Psi^{-}$ 
will destabilize their Fermi surface and engender a condensate 
$\langle\Psi^+\Psi^-\rangle\ne0$ that breaks $\U(1)_B$ symmetry 
spontaneously. (Note that $\Psi^+\Psi^-$ is charge neutral and 
does not break $\U(1)$ gauge symmetry.) 

In Phase III the long-distance theory comprises  
a massless photon and neutral fermions $\Psi^{\rmA=3}$ at finite $\mu$. 
It naively appears as though $\U(1)_B$ symmetry is unbroken,  
but as will be shown in Sec.~\ref{sc:u1break}, it could be spontaneously 
broken through an effective interaction mediated by heavy charged particles. 
Recalling the intuitive pictures for Phase I and II above, we conjecture 
that actually $\U(1)_B$ is broken \emph{everywhere} in the phase diagram 
(Fig.~\ref{fg:quark_num_density_Nc2}).%
\footnote{This is a hypothesis at exactly zero 
temperature. As soon as the imaginary-time direction is compactified, 
the true symmetry breaking would be replaced with a quasi-long-range order 
\cite{Mermin:1966fe,*Hohenberg:1967zz,*Coleman:1973ci}.}

\subsection{\label{sc:bion}Monopoles, bions and semiclassical confinement}

Let us focus on Phase III for $\Nc=2$ and $\NfD=1$. We assume $m=0$ 
for simplicity and switch to the two-component notation 
$\Psi^{\rmA=3}=L^{-1/2}\bep\psi_1 \\ \sigma^2\psi_2^{*}\eep$, 
thereby ensuring that $\psi_{1,2}$ possess the canonical dimension of spinors 
in three dimensions. 
The tree-level effective theory at small $L$ reads 
\ba
	S & = \!\!\int \!\! \dd^3x \bigg[\frac{1}{4g_3^2}F_{ij}^2 
	+ \psi_1^\dagger (i \sigma^i \der_i - \mu) \psi^{}_1 
	+ \psi_2^\dagger (i \sigma^i \der_i + \mu) \psi^{}_2
	\bigg] ,
	\label{eq:treeEFT}
\ea
where $\sigma^{1,2,4}\equiv (\sigma^1, \sigma^2, -i \1_2)$ and $g_3$ is defined in (\ref{g3}). 
Notice that $\mu\ne 0$ breaks the global $\SU(2)_f$ symmetry down to $\U(1)_B$ that 
rotates $\psi_1\to \ee^{i\varphi}\psi_1$ and $\psi_2\to \ee^{-i\varphi}\psi_2$. 

At a nonperturbative level, we also have instanton-monopoles, 
reflecting the compact nature of the $\U(1)$ gauge group. 
At first sight, Polyakov's mechanism of 
confinement due to Debye screening by monopoles \cite{Polyakov:1976fu} 
seems to apply, but this is not the case here: monopoles are accompanied by 
fermionic zero modes%
\footnote{For the index theorem at finite $\mu$, see 
Refs.~\cite{Gavai:2009vb,Kanazawa:2011tt,Bruckmann:2013rpa}.} 
and do not generate a mass gap for photons 
\cite{Affleck:1982as}.%
\footnote{The fermionic zero modes of monopoles are absent if one starts from a genuine 
compact $\U(1)$ theory rather than breaking a non-Abelian group with a Higgs 
mechanism. In such a case the dynamics of monopoles at long distances 
can be different from that in $\RR^3\times S^1$; see e.g., 
Refs.~\cite{Ioffe:1989zz,*Marston:1990bj,*Murthy1991,*Kleinert:2002uv,
*Herbut:2003bs,*Herbut2003PRB,*Hermele:2004hkd}.}%
\footnote{When $m\ne0$, fermionic zero modes can be soaked up by the mass term and 
monopoles are allowed to yield a bosonic potential \cite{Unsal:2010qh,Poppitz:2012sw}. 
Similarly, if the difermion condensate $\langle\psi\psi\rangle$ forms dynamically, it would 
lift the fermionic zero modes and induce a monopole-induced potential.  
As we will see in Sec.~\ref{sc:u1break}, indeed such a condensate can form, but it occurs
only at a super-soft scale and can be safely ignored at the scale of average  
monopole-monopole separation $\sim L\exp(S_m/3)$.} 
The effective theory at length scales $\gg L/g$ gains an additional piece 
that accounts for this effect \cite{Unsal:2007vu,*Unsal:2007jx,Argyres:2012ka}, 
\ba
	\delta S = G \int \!\! \dd^3x \;\Big[
		\cos \chi \cdot \det\limits_{1\leq I, J \leq 2\NfD}\;
		(\psi_{I}^{\rm T} \sigma^2 \psi^{}_{J}) + \text{h.c.}
	\Big]
	\label{eq:detterm}
\ea
where the scalar field $\chi(x)$ is a dual photon with $2\pi$-periodicity, 
related to the original field via Abelian duality relation as  $F_{ij}\sim \epsilon_{ijk}\der_k \chi$. 
The  monopole operator ('t~Hooft  type vertex for the monopole) \eqref{eq:detterm} evidently 
respects the $\SU(2)_f$ flavor symmetry. One can also check 
that \eqref{eq:detterm} is invariant under the  anomaly free discrete subgroup $(\ZZ_8)_A$, which acts as  
$\psi\to\ee^{i2\pi/8}\psi$, and  $\chi\to \chi+\pi$.
The factor $G$ can be extracted from the monopole measure, see  Sec.~4 of 
Ref.~\cite{Argyres:2012ka}.%
\footnote{Normalization of the fermion kinetic term in \eqref{eq:treeEFT} is 
different from that in Ref.~\cite{Argyres:2012ka}.} 
Up to an $\calO(1)$ numerical factor,
\ba
	\begin{split}
		G & \sim g^{-4} \ee^{-4\pi^2/g^2} L^{4\NfD-3}
		\\
		& = g^{-4} \ee^{-4\pi^2/g^2} L \qquad 
		\text{for}~\NfD=1\,,
	\end{split}
	\label{eq:GdepL}
\ea
where $g=g(L^{-1})$ is the renormalized coupling at the scale $L^{-1}$.

\subsubsection*{Magnetic bion formation at $\mu \neq 0$}
In order to see mass gap generation for photons, one needs to 
go to the next order in the semiclassical expansion. The point is that 
due to the compactness of the $A_3$ holonomy there is an additional class 
of monopole-instanton (called KK monopoles) in $\RR^3\times S^1$ \cite{Lee:1997vp,Lee:1998bb,Kraan:1998pm,*Kraan:1998sn}. 

There exists a topologically neutral, but magnetically charged  combination of the BPS-monopole with KK-anti-monopole  
called \emph{magnetic bion} \cite{Unsal:2007vu,*Unsal:2007jx}, which induces the operator $(\cos 2 \chi)$. 
Note that this operator is invariant under the action of  discrete chiral symmetry $\chi\to \chi+\pi$.
In some respect, this phenomenon is similar to the formation of 
instanton--anti-instanton molecules in thermal QCD \cite{Ilgenfritz:1994nt,Schafer:1994nv}, and in others it differs from it as 
the magnetic bion has still a topological quantum number for its magnetic charge. 
The physics of multi-instanton correlation in QCD with chemical potential has been studied in, e.g.,
Refs.~\cite{Carter:1998ji,Schafer:1998up,Rapp:1999qa}. 
In quark matter at high density, instantons of large sizes are exponentially suppressed 
due to the Debye screening of color-electric fields inside instantons \cite{Shuryak:1982hk}. 
In contrast, this is of no concern here because the medium of 
$\Psi^{{\rmA}=3}$ is transparent for monopoles. 

Below, we would like to explicitly calculate the effect of chemical potential $\mu$ on the magnetic bion formation. 
The operator ${\cal B}(x)$ of a magnetic bion 
involves a connected correlator of two monopole 
operators \cite{Unsal:2007vu,*Unsal:2007jx,Argyres:2012ka}: 
\ba
	\! \mathcal{B}(x) \sim G^2 \ee^{2i\chi(x)}
	\!\!\!\! \int \!\!\! \dd^3y\, 
	\ee^{-V_{\rm C}(y)} 
	\Big\langle
		\! \det \! \big(\psi\psi(y)\big) \det \! \big(\psi^\dagger\psi^\dagger(0)\big)
	\Big\rangle
	\label{eq:bionamp}
\ea
with the repulsive Coulomb potential
\ba
	V_{\rm C} (y) \equiv \frac{4\pi}{g_3^2}\frac{1}{|y|}\,.
\ea
At distances $\gg L$ the correlator of 't~Hooft vertices in \eqref{eq:bionamp} can be 
computed with a free fermion propagator 
$S_{\rm F}(x;\mu)=\langle \psi^{}_1(x)\psi_1^\dagger(0)\rangle$ 
and $S_{\rm F}(x;-\mu)=\langle \psi^{}_2(x)\psi_2^\dagger(0)\rangle$, 
resulting in
\ba
	& \Big\langle
		\! \det \! \big(\psi\psi(x)\big) \det \! \big(\psi^\dagger\psi^\dagger(0)\big)
	\Big\rangle  
	\notag
	\\
	\propto ~ & \tr \big[ S_{\rm F}^{\mathstrut}(x;\mu)
	S_{\rm F}^\dagger(x;\mu) \big] 
	\tr \big[ S_{\rm F}^{\mathstrut}(x;-\mu)	S_{\rm F}^\dagger(x;-\mu) \big] 
	\notag \\
	\propto~& \big\{(\der_1 D_+)^2 + (\der_2 D_+)^2 
	+ [(\der_4-\mu)D_+]^2 \big\}
	\notag
	\\
	& \times 
	\big\{(\der_1 D_-)^2 + (\der_2 D_-)^2 
	+ [(\der_4 + \mu)D_-]^2 \big\}\,,
	\label{eq:detcor}
\ea
where $D_{\pm}(x)\equiv D(x;\pm \mu)$ is a bosonic propagator 
that solves the Klein-Gordon-type equation
\ba
	[ -\der_1^2 - \der_2^2 - (\der_4-\mu)^2 ] D(x;\mu) = \delta(x)\,.
\ea
The explicit form of $D(x;\mu)$ and its properties are summarized in Appendix \ref{ap:propagator}. In the above we used \eqref{eq:SDap} there.  
At $\mu=0$, $D\propto |x|^{-1}$ and the correlator \eqref{eq:detcor} is just 
proportional to $|x|^{-8}$. However, once $\mu \neq 0$, there is anisotropy in space ($x^{1,2}$) 
and imaginary time ($x^4$); the weight \eqref{eq:detcor} exhibits a Friedel-type oscillation 
in spatial directions, reflecting the presence of a sharp Fermi surface, 
which tends to render bions anisotropic. 

Recalling that the three-dimensional Coulomb interaction is governed by the dimensionful coupling 
$g_3^2\equiv g^2/L$, it naturally follows that the strength of the bion deformation is measured by the ratio 
\ba
	\wt{\mu} \equiv \frac{\mu}{g_3^2} = \frac{\mu L}{g^2} \,.
\ea
If we fix $\mu$ and let $L\to 0$, then $\wt{\mu}\sim \mu L \log \frac{1}{L}\to 0$,  
i.e., the effect of $\mu$ on bions inevitably disappears. 
In order to see a nontrivial deformation of bions one 
has to increase $\mu$ in the course of compactification so that $\mu\gtrsim\frac{1}{L \log \frac{1}{L}}$. 
At the same time, the condition $\mu<\frac{1.98}{L}$ must also be satisfied,  
in order to stay inside Phase III (cf.~Fig.~\ref{fg:potential_Nc2_Nf1_m0}).

To simplify the $\wt{\mu}$-dependence of the weight 
we make all variables dimensionless in units of $g_3^2$: 
\ba
	\wt{y}& \equiv g_3^2 y \,, 
	\\
	\rho & \equiv \sqrt{\wt{y}_1^2+\wt{y}_2^2}\,,
	\\
	\tau & \equiv \wt{y}_4 \,,
	\\
	\wt{D}_{\pm}(\rho,\tau)& \equiv g_3^{-2}D(y;\pm\mu)\,. 
\ea
Using \eqref{eq:GdepL} we find that the bion amplitude is given by
\ba
	\mathcal{B}(x) \sim L^{-3}g^2 \ee^{-8\pi^2/g^2}
	W(\wt{\mu}) \ee^{2i \chi(x)}
	\label{eq:Brep}
\ea
with a dimensionless function $W$ defined by
\ba
	W(\wt{\mu}) & \equiv 
	\int_{0}^{\infty} \!\!\! \dd \rho\,\rho \! 
	\int_{-\infty}^{\infty}\!\!\! \dd \tau\;w(\rho,\tau;\wt{\mu})\,,
	\\
	w(\rho,\tau;\wt{\mu}) & \equiv \ee^{-4\pi/\sqrt{\rho^2+\tau^2}}
	\big\{(\der_\rho \wt{D}_+)^2 + [(\der_{\tau} - \wt{\mu})\wt{D}_+]^2 \big\}
	\notag
	\\
	& \quad \times 
	\big\{(\der_\rho \wt{D}_-)^2 + [(\der_{\tau} + \wt{\mu})\wt{D}_-]^2 \big\}
	\,.
	\label{eq:wdef}
\ea
The prefactors in \eqref{eq:Brep} are in precise 
agreement with Refs.~\cite{Anber:2011de,Argyres:2012ka} 
for the case of two Weyl fermions. 

%%%%%%%%%%%%%%%%%%%%%%%%%%%%%%%%%%%
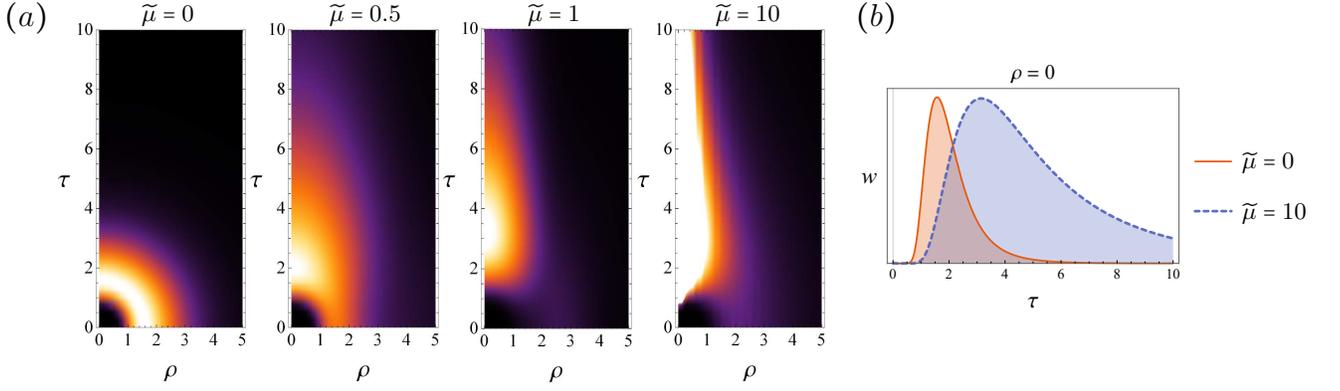
\begin{figure*}[bt]
	\hspace{-13pt}
	\includegraphics[width=.57\textwidth]{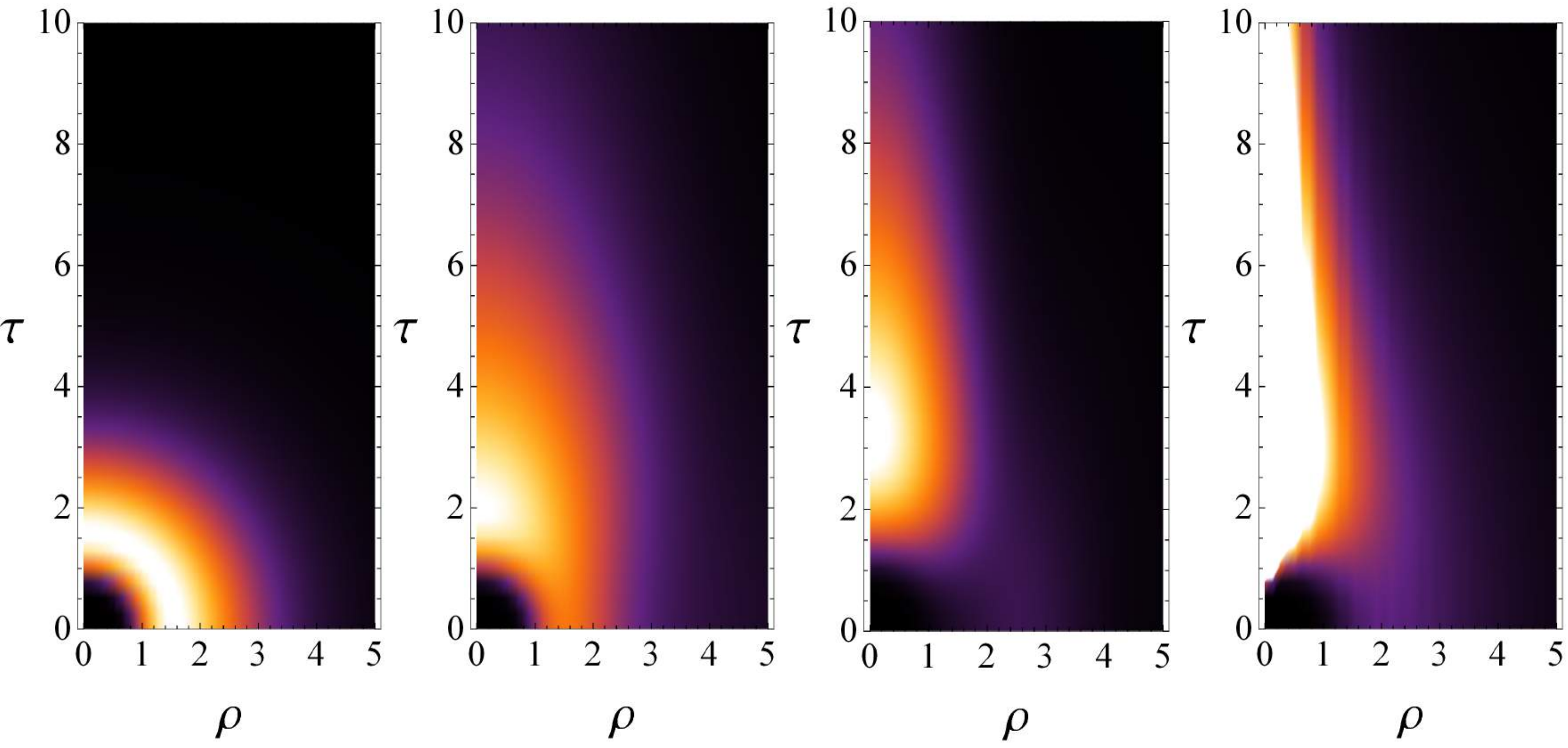}
	\put(-311,136){\large $(a)$}
	\put(-259,138){$\wt{\mu}=0$}
	\put(-188,138){$\wt{\mu}=0.5$}
	\put(-113,138){$\wt{\mu}=1$}
	\put(-42,138){$\wt{\mu}=10$}
	\quad 
	\raisebox{10mm}{
	\includegraphics[width=.28\textwidth]{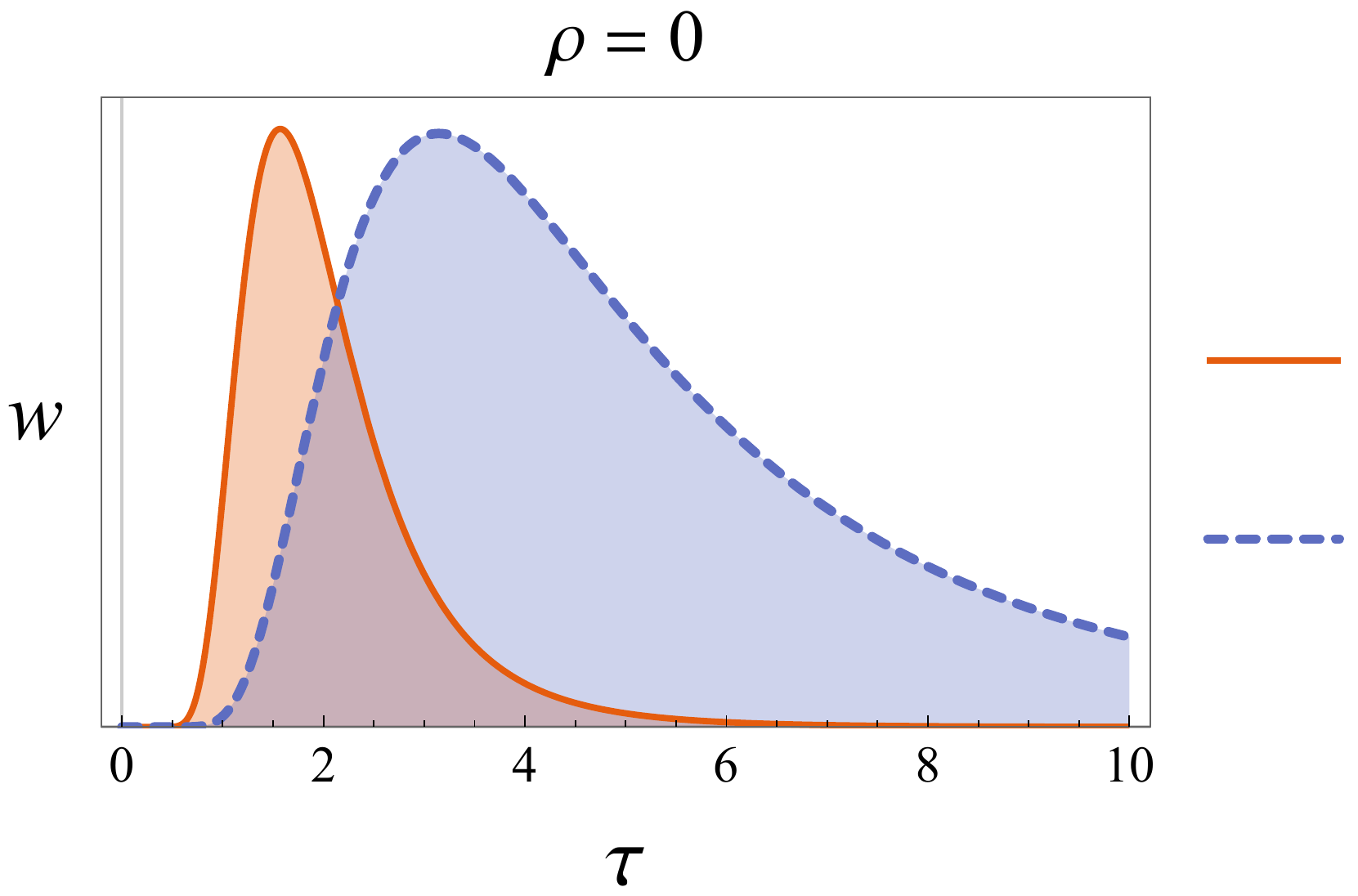}}
	\put(-144,136){\large $(b)$}
	\put(2,82){$\wt{\mu}=0$}
	\put(2,63){$\wt{\mu}=10$}
	\caption{\label{fg:bion_w}%
	(a) The probability density $w(\rho,\tau;\wt{\mu})$ of a magnetic bion \eqref{eq:wdef}. 
	Brighter regions have larger $w$. (b) Comparison of $w$ at $\rho=0$ between $\wt{\mu}=0$ 
	and $\wt{\mu} =10$. The heights are made equal for better visibility. 
	}
\end{figure*}
%%%%%%%%%%%%%%%%%%%%%%%%%%%%%%%%%%%
All the $\wt{\mu}$-dependence is now encapsulated in $w$.  
As displayed in Fig.~\ref{fg:bion_w}(a), $w$ dramatically changes 
its profile as a function of $\wt{\mu}$. 
At $\wt{\mu}=0$, $w$ is isotropic with a sharp peak along the circle 
$\sqrt{\rho^2+\tau^2}=\pi/2\simeq 1.57$. 
This implies that the typical size of a magnetic bion is $\pi/(2g_3^2)=\pi L/(2g^2)$, 
as noted in Refs.~\cite{Anber:2011de,Argyres:2012ka}.  
As $\wt{\mu}$ grows, the weight of $w$ gradually moves to higher 
$\tau$, with a peak position at $\tau\sim 3$. 
At $\wt{\mu}=10$, $w$ is further focused on the $\tau$ axis 
with a quite narrow range of $\rho\ll 1$. This means that the magnetic bion 
tends to be rigidly oriented in the $x^4$ direction. 
One can also see a ripple-like pattern in the rightmost plot of Fig.~\ref{fg:bion_w}(a), 
caused by a Friedel oscillation of $\wt{D}$. 

Our numerical analysis indicates that the global maximum of $w$ 
is always located on the $\tau$ axis, so let us examine the 
behavior of $w$ at $\rho=0$ more closely. Using $\der_{\rho}\wt{D}\to 0$ 
as $\rho\to 0$, one gets $w(0,\tau;\wt{\mu}) = 
\ee^{-4\pi/\tau}[(\der_{\tau} - \wt{\mu})\wt{D}_+]^2
[(\der_{\tau} + \wt{\mu})\wt{D}_-]^2$. Substituting the analytic form 
\eqref{eq:Dr0} of $\wt{D}$ in Appendix~\ref{ap:propagator} one obtains
\ba
	w(0,\tau;\wt{\mu}) = \frac{\ee^{-4\pi/\tau}}{(4\pi)^4}
	\frac{(1+\wt{\mu}\tau)^2
	(-1+ \wt{\mu}\tau+2\ee^{-\wt{\mu}\tau})^2}{\tau^8} 
\ea
for $\tau>0$. It shows that the peak of $w$ evolves from $\tau=\pi/2$ 
at $\wt{\mu}=0$ to $\tau = \pi$ at $\wt{\mu}\gg 1$. This behavior is 
illustrated in Fig.~\ref{fg:bion_w}(b). 

In summary, we have found two features of magnetic bions at  
$\mu\gtrsim\frac{1}{L\log \frac{1}{L}}$. First, the monopole--anti-monopole 
pair inside a bion is predominantly oriented in the imaginary-time direction. 
This is due to the fact that fermions with $\mu$ favor hopping in the temporal 
direction, which has been observed in the context of instantons in dense matter as well
too \cite{Schafer:1998up,Rapp:1999qa}.    
Secondly, the bion has a typical size $\pi/g_3^2=\pi L/g^2$. This is 
twice larger than that at $\mu=0$.

\subsubsection*{Semiclassical confinement at $\mu \neq 0$}
If we replace the integral of $w$ with its peak value, 
we obtain a crude estimate
\ba
	W(\wt{\mu}) \propto \wt{\mu}^4 \qquad \text{at}\quad \wt{\mu}\gg 1\,.
	\label{eq:Wasy}
\ea
It has an interesting implication for the dual photon mass. Let us recall that 
in addition to \eqref{eq:Brep} there are also operators associated with anti-bions. 
Their sum reads 
\ba
	\mathcal{B}+\mathcal{B}^\dagger \sim L^{-3}g^2 \ee^{-8\pi^2/g^2}
	W(\wt{\mu}) \cos(2\chi) \,.
	\label{eq:Bpote}
\ea
This potential has minima at $\chi=0$ and $\pi$ associated with the spontaneous breaking of discrete 
chiral symmetry. (Note that an order parameter for the discrete chiral symmetry is $e^{i \chi}$ and the two vacua correspond to 
 $\langle e^{i \chi} \rangle = \pm 1$.)
At these points 
the dual photon $\chi$ acquires a nonperturbative mass gap, given by
\ba
	M_{\chi} \sim \frac{1}{L}\ee^{-4\pi^2/g^2}\sqrt{W(\wt{\mu})}\,.
\ea
This means that the perturbatively massless photon is Debye-screened 
by the plasma of magnetic bions \cite{Unsal:2007vu,*Unsal:2007jx}. 
The fact that $W(\wt{\mu})$ asymptotically increases with $\wt{\mu}$ implies that 
$\mu$ tends to \emph{enhance} nonperturbative effect on the gauge field.  

Using \eqref{eq:Wasy} and the one-loop $\beta$ function for $g$, 
\ba
	\label{beta}
	e^{-4\pi^2/g^2} \sim (L \Lambda)^{7/3},
\ea 
one finds
\ba
	\frac{M_{\chi}}{\Lambda} & \sim (L\Lambda)^{4/3}\wt{\mu}^2 
	\notag \\
	& \sim \Big(\log \frac{1}{L\Lambda}\Big)^2 (L\Lambda)^{10/3}  
	\Big(\frac{\mu}{\Lambda}\Big)^2\,,
	\label{eq:M34}
\ea
with a renormalization-group invariant scale $\Lambda$.
The mass gap for the dual photon implies that the fundamental Wilson loop 
in $\RR^3$ obeys area law; quarks are permanently confined. 
The string tension is given by $T \sim \frac{g^2}{L}M_{\chi}$  
\cite{Polyakov:1976fu}. The $L$ dependence follows from 
\eqref{beta} and \eqref{eq:M34} as
\ba
	\frac{T}{\Lambda^2} \sim 
	\Big(\log \frac{1}{L\Lambda}\Big) (L\Lambda)^{7/3}  
	\Big(\frac{\mu}{\Lambda}\Big)^2\,.
\ea
This estimate is valid for $\mu$ in the range
\ba
	\frac{1}{L\log\frac{1}{L\Lambda}} \ll \mu 
	< \frac{1.98}{L}\,.
	\label{eq:mucond}
\ea

\subsection{\label{sc:u1break}Superfluidity}

Generically a fermionic system at finite density can be unstable toward 
pair condensation if there is an attractive channel in their interaction. 
It is therefore important to ask what interactions can happen for 
$\psi_{1,2}$ in the effective theory \eqref{eq:treeEFT}. 
Although they do not couple to photons by themselves, they interact through 
\emph{two} mechanisms of different physical origins. 
The first one is mediated by monopole-instantons. 
As was shown in the previous section 
the dual photons are screened at distances beyond 
$1/M_{\chi}$. As $\chi$ settles in one of the two minima $\{0,\pi\}$ 
of the potential \eqref{eq:Bpote},%
\footnote{This breaks the anomaly-free axial $\ZZ_{8}$ down to $\ZZ_4$ 
spontaneously.} one obtains 
from \eqref{eq:detterm} a four-fermion operator 
\ba
	\pm G\, \Big[ \det_{I,J}\;(\psi_{I}^{\rm T} 
	\sigma^2 \psi^{}_{J}) + \text{h.c.} \Big] \,. 
	\label{eq:monin}
\ea
Note that the 4d instanton would induce an 8-fermion operator, with much
weaker coefficient (of order $G^2$).
 In this sense, the monopole-instantons 
are much more important than 4d instantons. 

\begin{figure}[b]
	\centering
	\includegraphics[width=.45\textwidth]{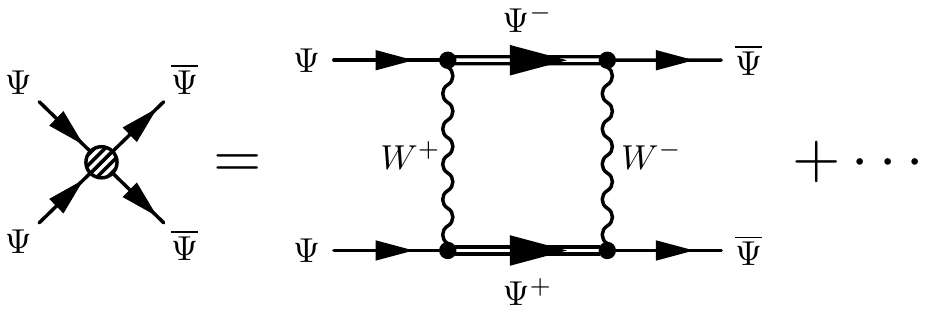}
	\vspace{-4pt}
	\caption{\label{fg:box_dgm}%
	A four-fermion operator that arises from integrating out 
	heavy charged particles. 
	}
\end{figure}

Yet another kind of interaction stems from integrating out 
heavy $\Psi^{\pm}$ and $W^{\pm}$ that are charged under $\U(1)$.  
As an example, we show in Fig.~\ref{fg:box_dgm} an effective 
interaction in the diquark channel where 
the 4-component Dirac spinor in the Cartan subalgebra 
$\Psi^{\rmA=3}$ is notated by $\Psi$. 
%%%%%%%%%%%%%%%%%%%%%%%%%%%%%%%%%%%
%%%%%%%%%%%%%%%%%%%%%%%%%%%%%%%%%%%
There are a myriad of operators that can be perturbatively 
generated this way, e.g., $(\bar\Psi \gamma^3\Psi)^2$, 
$(\bar\Psi \gamma^i\Psi)^2$ and 
$(\bar\Psi \gamma^3 \gamma^i \gamma^j\Psi)^2$,  
to name but a few. They are all invariant under $\U(1)_A$ and hence 
do not couple to the dual photon.  Here we do not systematically 
enumerate all possible forms of the interaction, but instead try to capture 
an essential physical outcome by considering the minimal interaction 
in the (pseudo-)scalar channel 
\ba
	- H \big[ (\bar\Psi\Psi)^2 + (\bar\Psi i \gamma_5 \Psi)^2 
	+ |\Psi^{\rm T}C\Psi|^2 + |\Psi^{\rm T}C\gamma_5\Psi|^2 
	\big]
	\label{eq:nf4}
\ea
with a charge-conjugation matrix $C=\diag(-\sigma^2,\sigma^2)$ 
and $\gamma_5=\diag(\1_2,-\1_2)$. This interaction, being perturbatively induced,  
preserves $\U(1)_A\times \SU(2)_f$ \cite{Zhang:2010kn}. 
The coupling $H$ should be 
$\propto L$ according to dimensional analysis, 
and it must be multiplied by $g^4$ because there 
are at least four bare vertices in the process 
of Fig.~\ref{fg:box_dgm}. Thus
\ba
	H \propto g^4 L\,. \label{eq:Hestm}
\ea
Since $G$ [cf.~\eqref{eq:GdepL}] is proportional to  
$\ee^{-4\pi^2/g^2}\ll 1$, 
\eqref{eq:Hestm} means that $|H|\gg G$.  

The sign of $H$ is not fixed by symmetries alone. 
In this regard we would like to note that the process of Fig.~\ref{fg:box_dgm} 
is of the same kind as the Van der Waals interaction in QED, which originates 
from the two-photon exchange and is attractive because it arises 
at second order in perturbation theory \cite{Landau:1991wop}.  
Similar long-range force between color-singlet hadrons in QCD, 
originating from the two-gluon exchange process, is also known to be 
attractive \cite{Fishbane:1977ay,Appelquist:1978rt,Fujii:1999xn}.  
Therefore in the following analysis we will assume $H>0$ and pursue its 
physical consequences, leaving a microscopic verification 
to future work. We remark that 
the phase diagram shown earlier as Fig.~\ref{fg:pd} was also based on the assumption 
that $H>0$.

\subsubsection*{Fermionic effective theory}
Incorporating both types of interactions [\eqref{eq:monin} and \eqref{eq:nf4}], and dropping 
the quark-antiquark vertices because they are not relevant to 
the BCS instability, we end up with the fermionic effective theory  
\ba
	S & = \int \dd^3x \, \Big\{  
	\bar{\Psi}(\gamma_i\der_i - \mu\gamma_4)\Psi
	\pm G\, \Big[ \det_{I,J}\;(\psi_{I}^{\rm T} 
	\sigma^2 \psi^{}_{J}) + \text{h.c.} \Big] 
	\notag
	\\
	& \quad -H \, \big( 
		|\Psi^{\rm T}C\Psi|^2 + |\Psi^{\rm T}C\gamma_5\Psi|^2
	\big)
	\Big\} \,,
	\label{eq:Ss}
\ea
where the sign in front of $G$ depends on whether 
$\langle \chi \rangle=0$ or $\pi$.   
Analogous four-fermion models in three and four dimensions have been studied before 
\cite{Nambu:1961tp,*Nambu:1961fr,Rosenstein:1988dj,*Rosenstein:1990nm}. We note that 
while the models in Refs.~\cite{Nambu:1961tp,*Nambu:1961fr,Rosenstein:1988dj,*Rosenstein:1990nm} 
were introduced on phenomenological grounds, 
\eqref{eq:Ss} is a faithful description of  the microscopic dynamics of adjoint QCD in the 
circle-compactification limit.

A remark on the parity of the ground sate is in order. 
Because the second line of \eqref{eq:Ss} does \emph{not} 
distinguish between $0^+$ and $0^-$ diquarks, 
one shall look at the monopole-induced operator.  
With the help of the relation 
\ba
	\!\!
	\det_{I,J}(\psi_I^{\rm T}\sigma^2\psi^{}_J) + \text{h.c.} 
	\notag & = \frac{3}{2} [(\psi_1^{\rm T}\sigma^2\psi^{}_1)
	(\psi_2^{\rm T}\sigma^2\psi^{}_2) + \text{h.c.}]
	\\
	& = \frac{3}{4}
	( |\Psi^{\rm T}C\Psi|^2 - |\Psi^{\rm T}C\gamma_5\Psi|^2 ), 
	\label{eq:er50t}
\ea
we observe that interaction in the $0^+$ channel is stronger 
than $0^-$ for $\langle\chi\rangle=0$. 
However the opposite happens in the other vacuum 
$\langle\chi\rangle=\pi$. Thus, after all, the effective theory \eqref{eq:Ss} 
\emph{per se} does not uniquely determine the parity of the 
ground state, and prompts us to add a small but 
nonzero external source field to resolve the ambiguity. 
If one inserts to the action a real mass term $m\bar\Psi\Psi$ 
with an arbitrarily small $m$, the ambiguity is resolved.%
\footnote{In QCD-like theories at finite density 
\cite{Kogut:1999iv,Son:2000xc,Kogut:2000ek}, 
the parity of the diquark or pion condensate 
was automatically determined because the parity of the ground 
state at zero chemical potential was fixed by a real mass term.} 
There are two ways to check this. First,  
one can invoke QCD inequalities generalized to $\mu\ne0$ 
\cite{Kogut:1999iv,Cherman:2011mh,Kanazawa:2011tt} 
to argue that $\Psi^{\rm T}C\gamma_5 \Psi$ is the lightest bilinear 
field.%
\footnote{This observation lends support to our strategy to 
consider a simplified effective theory \eqref{eq:nf4} 
from the very beginning.} 
We stress that this argument works equally in both $\RR^4$ 
and $\RR^3\times S^1$. The second argument to check parity only works 
in the semiclassical small-$L$ regime. Namely, when the above sources are 
present in the action, they will absorb the fermionic zero modes inhabiting 
the monopoles and allow for a bosonic potential 
$\sim m^2 \cos \chi$ to be generated 
on top of the bion-induced potential \eqref{eq:Bpote}.  
This potential, however small in magnitude, can and do 
lift the degeneracy between $\chi=0$ and $\pi$, 
and consequently fix the parity of the ground state.  

From the consideration above, one may assume that 
the interaction is stronger for the $0^+$ channel than for $0^-$, 
tacitly assuming the presence of a suitable 
infinitesimal external field in the action. Now, 
since the monopole-induced interaction is nonperturbatively 
small, it can be safely dropped, and we arrive at a simpler theory 
that sufficiently serves our purpose of probing the 
ground state of the fermionic sector: 
\ba
	S & = \!\! \int \!\! \dd^3x \, \big[
	\bar{\Psi}(\gamma_i\der_i - \mu\gamma_4)\Psi
	- H |\Psi^{\rm T}C\gamma_5\Psi|^2 \big]\,. 
	\label{eq:S4c}
\ea

\subsubsection*{Gap equation}
We solve the theory in the mean-field approximation following 
the standard procedure \cite{Hatsuda:1994pi,Buballa:2003qv}. 
This is expected to be accurate, given the smallness of the 
coupling $H\mu \propto g^4 \muu \ll 1$.  
The Hubbard--Stratonovich transformation applied to 
\eqref{eq:S4c} with an auxiliary complex field $\Delta(x)$ yields 
\ba
	S & = \!\! \int \!\! \dd^3x \, \Big[
	\bar{\Psi}(\gamma_i \der_i - \mu\gamma_4)\Psi 
	+ \frac{|\Delta|^2}{4H}
	- \frac{1}{2}(\Delta^* \Psi^{\rm T}C\gamma_5\Psi+\text{h.c.}) \Big].
\ea
With a suitable $\U(1)_B$ rotation, one can take $\Delta\geq 0$ 
without loss of generality. 
In the Nambu--Gor'kov basis $(\Psi, C\bar{\Psi}^{\rm T})$ we have 
for the energy density
\ba
	U(\Delta) & \equiv - \frac{1}{V_{\RR^3}}\log Z 
	\notag \\
	& = \frac{\Delta^2}{4H}
	- \frac{1}{2}\int \!\!\! \frac{\dd^3 p}{(2\pi)^3} \tr\log 
	\bep
		\Delta \gamma_5 & i\slashed{p}+\mu\gamma_4
		\\
		i \slashed{p} - \mu\gamma_4 & - \Delta \gamma_5
	\eep
	\notag \\
	& = \frac{\Delta^2}{4H} - \int \!\!\! \frac{\dd^3 p}{(2\pi)^3}
	\Big\{ \log[ p_4^2+(|\bm{p}_{\perp}|-\mu)^2+\Delta^2 ] 
	\notag
	\\
	& \qquad + \log[ p_4^2+(|\bm{p}_{\perp}|+\mu)^2+\Delta^2 ] 
	\Big\}
	\notag \\
	& = \frac{\Delta^2}{4H} - \int' \! \frac{\dd p_1\dd p_2}{(2\pi)^2}
	\Big\{ \sqrt{(|\bm{p}_{\perp}|-\mu)^2+\Delta^2} 
	\notag
	\\
	& \qquad 
	+ \sqrt{(|\bm{p}_{\perp}|+\mu)^2+\Delta^2} \Big\}\,.
\ea
The final integral with a prime is to be regularized with a cutoff. 
Then the gap equation $\der U(\Delta)/\der \Delta=0$ for a nontrivial solution 
$\Delta\ne 0$ is given by
\ba
	\frac{1}{H} = \!\! \int' \!\frac{\dd p\;p}{\pi} 
	\bigg(\frac{1}{\sqrt{(p-\mu)^2+\Delta^2}}
	+\frac{1}{\sqrt{(p+\mu)^2+\Delta^2}}\bigg)\,.
\ea
As $\Delta\to 0$ the first term yields IR divergence at the Fermi surface. 
Retaining only the first term, and imposing a momentum 
cutoff $\Lambda'\,(>\mu)$, one obtains
\ba
	\frac{1}{H} \simeq \frac{\mu}{\pi} 
	\log \frac{4\mu(\Lambda'-\mu)}{\Delta^2}\,,
\ea
or
\ba
	\Delta \propto \sqrt{\mu(\Lambda'-\mu)} \exp
	\mkakko{ -\frac{\pi}{2H\mu} } .
	\label{eq:gapsol}
\ea
Physically, this means that fermions acquire an 
energy gap $\Delta$ and 
the $\U(1)_B$ symmetry is spontaneously broken by the condensate
$\langle\Psi^{\rm T}C\gamma_5\Psi\rangle\ne 0$;
the system is in a superfluid phase.%
\footnote{Since $\Psi$ does not couple to the $\U(1)$ photon, 
this phase is not a superconductor.} 
We stress that the parity of the diquark condensate is the same 
at small $L$ and large $L$ [recall our remarks below \eqref{eq:er50t}]. 
Note that in this analysis the density of fermions need not be high; the computation 
in this section is valid for any $\mu>0$ as long as we are in Phase III ($\mu<1.98/L$), 
which means that superfluidity may take place at \emph{arbitrarily small} $\mu$. 
This is a remarkable result. 

Plugging \eqref{eq:Hestm} and $\Lambda'\sim L^{-1}$ 
into \eqref{eq:gapsol}, we gain the crude estimate
\ba
	\Delta \propto \frac{\sqrt{\muu}}{L}
	\exp\mkakko{-\frac{1}{g^4 \muu}}.
	\label{eq:gapest}
\ea
Notably, this ultrasoft scale is far smaller than 
the dual photon mass $M_\chi\propto L^{-1}
\exp(-4\pi^2/g^2)$ and is invisible 
at any finite order of the semiclassical expansion. 
Since the extent of the Cooper pair $1/\Delta$ far exceeds the typical 
bion size, the diquark condensate is not expected to modify 
the fermion-binding mechanism of bions in Sec.~\ref{sc:bion}.

The various length scales in Phase III may be summarized as follows: 
\ba
	\scalebox{0.9}{$\displaystyle
	\begin{array}{ccccccccccccc}
		r_{\rm m} & \lesssim & d_{\psi\text{-}\psi} & \!\ll\! & r_{\rm b} & \!\ll\!\! &  
		d_{\rm m\text{-}m} & \!\! \ll \!\!\!\! & d_{\rm b\text{-}b} & \!\!\!\! \ll \!\!\! & 
		\frac{1}{M_\chi} & \!\!\!\ll\!\!\!\!\!\!\! & \frac{1}{\Delta}~~~~~
		\\
		\downarrow &&\downarrow &&\downarrow &&\downarrow &
		&\downarrow &&\downarrow && \hspace{-17pt}\downarrow 
		\vspace{2pt}\\
		L && \frac{1}{\mu} && \frac{L}{g^2} & 
		& L \ee^{S_{\rm m}/3} && L \ee^{2S_{\rm m}/3} && 
		L \ee^{S_{\rm m}} && \!\!\!\! 
		\frac{L}{\sqrt{\muu}} \exp\big(\case{1}{g^4\muu}\big)
	\end{array}
	$}
\ea
($r_{\rm m}$:~monopole size; $d_{\psi\text{-}\psi}$:~interquark distance; 
$r_{\rm b}$:~bion size; $d_{\rm m\text{-}m}$:~intermonopole distance; 
$d_{\rm b\text{-}b}$:~interbion distance; $1/M_{\chi}$:~inverse Debye screening length; 
$1/\Delta$:~Cooper-pair size; $S_{\rm m}=4\pi^2/g^2$: the monopole action).

\subsubsection*{Low-energy effective theory for superfluid phonons}
In the far-infrared limit at energies below $\Delta$, the effective degree of freedom is 
phonon, i.e., the Nambu--Goldstone mode associated with the baryon number. 
It emerges as a phase fluctuation of the condensate and can be introduced as 
$\Delta \to \Delta \ee^{2i\phi(x)}$.
The low-energy effective theory of phonons can be derived by expanding 
the functional determinant of $\Psi$ in powers of the derivative of $\phi$ 
(the so-called gradient expansion \cite{Eguchi:1976iz}).  
The result is
\ba
	\mathcal{L}_{\rm eff} = f^2 \Big[ (\der_4 \phi)^2+
	v^2 \big\{ (\der_1 \phi)^2 + (\der_2 \phi)^2 \big\} \Big] 
	\label{eq:Leffphonon}
\ea
with 
\ba
	f^2 = \frac{\mu}{2\pi}\,, \qquad v^2 = \frac{1}{2}\,. 
	\label{eq:fandv}
\ea
The factor of 2 in the denominator of $v^2$ reflects the dimensionality of space. 
The effective theory \eqref{eq:Leffphonon} can also be derived 
from the equation of state $P=\mu^3/(6\pi)$ on the basis of symmetry 
arguments along the lines of Ref.~\cite{Son:2002zn}.   
Equation \eqref{eq:fandv} should be contrasted with $f^2\sim\mu^2$ 
and $v^2=1/3$ for the Nambu--Goldstone modes 
in high-density quark matter in four dimensions 
\cite{Son:1999cm,*Son:2000tu,Beane:2000ms,Fukushima:2005gt}.

\section{\label{sc:con}Conclusion}

In this paper we extended the bion mechanism of confinement \cite{Unsal:2007vu,*Unsal:2007jx} 
to nonzero quark chemical potential $\mu$ in adjoint QCD with \emph{spatial} compactification. 
In the first part, we performed a perturbative analysis of Wilson line potential 
(whose result gives a realization of Hosotani mechanism) in the presence of  $\mu$ and revealed a rich phase diagram 
in the space of $\mu$ and $m$. In the second part we studied the $\mu$-dependence of 
semiclassical configurations (monopoles and their molecules called magnetic bions) in the center-symmetric phase. 
In addition to the Coulomb interaction, monopole-instantons  also talk to each other via a fermion zero mode exchange, which 
at $\mu\ne 0$ is modified due to the anisotropic hopping amplitude of 
fermions. Consequently, bions favor a temporal orientation and its amplitude grows with $\mu$, leading to 
a larger mass gap of photons and a greater string tension. Intriguingly, neutral massless fermions that are free in the 
leaing-order perturbation theory may exhibit novel superfluidity triggered by  the combination of 
perturbatively induced four-fermion  operators and nonperturbatively induced  monopole operators  ('t~Hooft vertex).   
The analysis in this paper remains valid at any $\mu \neq 0$ as long as the compactified direction 
$S^1$ is small enough. It would be interesting to examine 
the dimensional crossover from small to large $S^1$ (Fig.~\ref{fg:pd})  
in future lattice simulations, to clarify how the BEC-BCS crossover region in $\RR^4$ is 
connected to the exotic phase structure in $\RR^3\times S^1$ found in this work. 
Dimensional crossover of an interacting Fermi gas has been actively investigated 
in the condensed matter physics community \cite{Valla2002,*PhysRevA.67.031601,
*PhysRevLett.106.105304,*PhysRevA.84.053624,*PhysRevLett.108.045302,
*PhysRevA.88.023612,*PhysRevLett.112.045301,*Toniolo2017} 
and it is of great interest to see what new physics will emerge for the case of 
relativistic quark matter.

\begin{acknowledgments}
	We are grateful to A.~Cherman 
	and T.~Sulejmanpasic %and T.~Sch\"afer 
	for valuable comments on the manuscript. 
	T.~K. was supported by the RIKEN iTHES project. 
	N.~Y. was supported by JSPS KAKENHI Grant No.~16K17703 and MEXT-Supported Program for 
	the Strategic Research Foundation at Private Universities, ``Topological Science'' (Grant No.~S1511006). 
	 M.~\"U.  was supported by the DOE under Grants No.~DE-SC0013036.
\end{acknowledgments}

\appendix
\section{\label{ap:oneloopdet}One-loop effective potential}
Below we will outline the derivation of the formula \eqref{eq:Veff1}.  
Let us momentarily put the system in a box of linear extent 
$L_\perp\times L_\perp\times L\times \beta$ with $\beta$ the 
inverse temperature and $L_\perp$ the length in the $x^{1,2}$ directions.
The Euclidean Dirac operator in the background \eqref{eq:A3} reads
\ba
	\DD(\mu) & =\gamma_1 \der_1 + \gamma_2 \der_2 + 
	\gamma_3 D_3^\text{ad} + \gamma_4 (\der_4 -\mu)\,,
\ea
where $D_3^\text{ad}$ is the adjoint covariant derivative that acts on a 
matrix test field $v=(v_{\rma \rmb})$ as
\ba
	(D_3^{\rm ad}v)_{\rma \rmb} & \equiv (\der_3 v+i[A_3, v])_{\rma \rmb} 
	\\
	& = [\der_3 +i(a_{\rma} - a_{\rmb})] v_{\rma \rmb}\,.
\ea
Since each component of $v$ except for its trace 
can be regarded as independent, we get \cite{Gross:1980br}
\begin{widetext}
\ba
	\Gamma_q (\Omega;\mu) & = \frac{1}{2}\tr \log [ -\DD(\mu)^2 + m^2 ]
	\notag \\
	& = 2 \tr \log \left[
	- \der_1^2 - \der_2^2 - (D_3^{\rm ad})^2 - (\der_4-\mu)^2 + m^2 
	\right]
	\notag \\
	& = 2 L_\perp^2 \!\!\int\!\!
	\frac{\dd p_1 \dd p_2}{(2\pi)^2} \sum_{p_3}\sum_{p_4}
	\bigg(
		\sum_{\rma, \rmb=1}^{\Nc}\log\left[
			p_{\perp}^2+(p_3 + a_\rma - a_\rmb)^2+(p_4+i\mu)^2+m^2
		\right] 
		- \log\left[\bm{p}^2+(p_4+i\mu)^2+m^2 \right]
	\bigg)
	\notag \\
	& = L_\perp^2 \!\! \int \!\!
	\frac{\dd p_1 \dd p_2}{(2\pi)^2} \sum_{p_3}\sum_{p_4}
	\bigg\{
		\sum_{\rma,\rmb=1}^{\Nc} \! \big(
			\log\left[ p_4^2+({\cal E}_{\rma\rmb}+\mu)^2 \right] + 
			\log\left[ p_4^2+({\cal E}_{\rma\rmb}-\mu)^2 \right] 
		\big)
		- \log\left[ p_4^2+({\cal E}+\mu)^2 \right] - \log\left[ p_4^2+({\cal E}-\mu)^2 \right]
	\bigg\}
	\,,  
	\label{eq:Gans}
\ea
where $p_{\perp}^2\equiv p_1^2+p_2^2$, ~$\bm{p}^2\equiv p_\perp^2+p_3^2$, 
~$\displaystyle p_3 = \frac{2n\pi}{L} (n\in\ZZ)$,~
$\displaystyle p_4 = \frac{(2\ell+1)\pi}{\beta} (\ell\in\ZZ)$, 
~${\cal E}\equiv \sqrt{\bm{p}^2+m^2}$ and    
\ba
	{\cal E}_{\rma \rmb} & \equiv \sqrt{p_{\perp}^2 + 
	(p_3+a_\rma - a_\rmb)^2+m^2} \,.
\ea
The summation over $p_4$ can be taken with the 
standard formula \cite{Kapusta:2006pm}
\ba
	\sum_{\ell = -\infty}^{\infty}\log[(2\ell +1)^2\pi^2+\omega^2] 
	= \omega + 2 \log (1+\ee^{-\omega})\,,
\ea
where an $\omega$-independent divergence has been subtracted. 
We thus obtain
\ba
	\Gamma_q (\Omega;\mu) & = 
	2 \beta L_\perp^2 \int\frac{\dd p_1 \dd p_2}{(2\pi)^2} \sum_{p_3}
	\bigg\{
		\sum_{\rma, \rmb=1}^{\Nc}\bigg(
			{\cal E}_{\rma\rmb} + 
			\frac{1}{\beta}\log\kkakko{ 1+\ee^{-\beta({\cal E}_{\rma\rmb}+\mu)} }
			+ \frac{1}{\beta}\log\kkakko{ 1+\ee^{-\beta({\cal E}_{\rma\rmb}-\mu)} }
		\bigg) 
		\notag
		\\
		& \quad - {\cal E} - \frac{1}{\beta}\log\kkakko{1+\ee^{-\beta({\cal E}+\mu)}}
		- \frac{1}{\beta}\log\kkakko{1+\ee^{-\beta({\cal E}-\mu)}}
	\bigg\}
\ea
and 
\ba
	\delta V_{\rm F}(\Omega;\mu) 
	& = - \lim_{\beta\to\infty}\frac{1}{\beta L^2_{\perp}L}
	\ckakko{ \Gamma_q (\Omega;\mu) - \Gamma_q (\Omega;0) } 
	\notag \\
	& = - \frac{2}{L}\int\frac{\dd p_1 \dd p_2}{(2\pi)^2} \sum_{p_3}
	\bigg\{
		\sum_{\rma, \rmb=1}^{\Nc}(\mu-{\cal E}_{\rma\rmb})
		\theta(\mu-{\cal E}_{\rma\rmb}) 
		- (\mu-{\cal E})\theta(\mu-{\cal E}) 
	\bigg\}
	\qquad \text{for}\quad \mu\geq 0\,.
\ea
The remaining integral may be done with the formula
\ba
	\int\frac{\dd p_1 \dd p_2}{(2\pi)^2} 
	(\mu-\sqrt{p_\perp^2+X}\,)\theta(\mu-\sqrt{p_\perp^2+X}\,)
	= \frac{1}{4\pi}\left(  \frac{1}{3}\mu^3-\mu X+\frac{2}{3}X^{3/2}  \right) 
	\theta(\mu^2-X)
	\qquad \text{for  }X\geq 0\,,
\ea
with the result
\ba
	\delta V_{\rm F}(\Omega;\mu) = - \frac{1}{2\pi L}
	\sum_{p_3}
	\bigg\{
		\sum_{\rma, \rmb=1}^{\Nc} 
		\left(  \frac{1}{3}\mu^3-\mu X+\frac{2}{3}X^{3/2}  \right)\theta(\mu^2-X)
		\bigg|_{X=(p_3+a_{\rma}-a_{\rmb})^2+m^2}
		- \left(  \frac{1}{3}\mu^3-\mu Y+\frac{2}{3}Y^{3/2}  \right)\theta(\mu^2-Y)
		\bigg|_{Y=p_3^2+m^2}
	\bigg\}.
\ea
Substituting $\displaystyle p_3 = \frac{2n\pi}{L} (n\in\ZZ)$ into this equation, 
we finally arrive at \eqref{eq:Veff1}. 
\end{widetext}

\section{\label{ap:propagator}\boldmath Free propagators with $\mu\ne 0$ in three dimensions}

We solve for the Euclidean propagators in $2+1$ dimensions in the presence of $\mu$. 
The fermion propagator $S_{\rm F}$ solves the equation
\ba
	(i\sigma^1\der_1+i\sigma^2\der_2+\der_4-\mu) 
	S_{\rm F}(x;\mu)=\delta(x)\1_2 \,.
\ea
The boson propagator solves the Klein-Gordon equation
\ba
	[ -\der_1^2 - \der_2^2 - (\der_4-\mu)^2 ] D(x;\mu) = \delta(x)\,.
\ea
They are related through
\ba
	S_{\rm F}(x;\mu) & = (i\sigma^1\der_1+i\sigma^2\der_2-\der_4+\mu) D(x;\mu) \,.
	\label{eq:SDap}
\ea
It is easy to verify that
\ba
	D(\bm{x},x_4; \mu) = D(\bm{x}, -x_4; -\mu), 
	\qquad \bm{x}=(x_1,x_2) \,.
\ea
Thus one can assume $\mu>0$ without loss of generality. 
Writing $r\equiv \sqrt{x_1^2+x_2^2}$\,, we have
\begin{align*}
	D(x;\mu) & = \int \frac{\dd^3p}{(2\pi)^3}
	\frac{\ee^{ipx}}{p_1^2+p_2^2+(p_4+i\mu)^2}
	\\
	& = \int_{-\infty}^{\infty} \!\!\! \dd p_4  \! 
	\int_0^\infty \frac{\dd p_\perp\,p_\perp}{(2\pi)^3} 
	\int_0^{2\pi} \!\!\!  \dd \theta 
	\; \frac{\ee^{i p_\perp r \cos\theta+i p_4 x_4}}{p_\perp^2+(p_4+i\mu)^2}
	\\
	& = \int_{-\infty}^{\infty} \!\!\! \dd p_4 \,\ee^{i p_4 x_4} \! 
	\int_0^\infty \frac{\dd p_\perp\,p_\perp}{(2\pi)^2} 
	\; \frac{J_0(p_{\perp}r)}{p_\perp^2+(p_4+i\mu)^2}
	\\
	& = \frac{1}{(2\pi)^2}\int_{-\infty}^{\infty} \!\!\! \dd p_4 
	\,\ee^{i p_4 x_4} 
	K_0 \big( \sqrt{(p_4+i\mu)^2}\, r \big)\,.
\end{align*}
An alternative expression which has no singularity at $r=0$ can also be 
obtained by integrating over $p_4$ first, with the result
\ba
	\!\! 
	D(x;\mu) = \frac{\ee^{\mu x_4}}{4\pi}\bigg\{
		\frac{1}{\sqrt{r^2+x_4^2}} - \! \int_0^\mu \!\! 
		\dd p_\perp \, J_0(p_\perp r)\ee^{-x_4 p_\perp}
		\! 
	\bigg\}\,\!. \!\!\!\!
\ea
These formulas give access to limiting behaviors of $D$ in some cases of interest: 
\begin{itemize}
	\item For $\mu=0$,
	\ba
		D(x;0)=\frac{1}{4\pi \sqrt{r^2+x_4^2}}\,.
	\ea
	\item For $r=0$,
	\ba
		D(x;\mu) & = \frac{1}{4\pi x_4}
		\big\{ 1 - \theta(-x_4) 2\ee^{-\mu |x_4|} \big\}
		\label{eq:Dr0}
		\\
		& \approx \frac{1}{4\pi x_4}\qquad \text{for} \quad \mu|x_4| \gg 1 \,.
	\ea
	\item For $x_4=0$,
	\ba
		D(x;\mu) & = \frac{1}{4\pi r}\int^{\infty}_{\mu r} \!\!\!\dd x\,J_0(x)
		\\
		& \approx \frac{1}{\sqrt{\mu}} 
		\frac{\cos\big( \mu r+\frac{\pi}{4} \big)}{(2\pi r)^{3/2}}
		\qquad\text{for} \quad \mu r\gg 1.
	\ea 
	\item For $\mu r\gg 1$ with any $x_4$, 
	\ba
		D(x;\mu) \approx \frac{r \cos\big( \mu r+\frac{\pi}{4} \big) + 
		x_4 \sin\big( \mu r+\frac{\pi}{4} \big)}
		{(2\pi)^{3/2}(r^2+x_4^2)\sqrt{\mu r\,}} \,.
	\ea
\end{itemize}

\bibliography{draft_QCD_on_torus_v13.bbl}
%\bibliographystyle{apsrev4-1}
%\bibliography{ref_toroidal}
\end{document}